\documentclass[prd,rd,showpacs,twocolumn,superscriptaddress]{revtex4-1}
\usepackage{amsfonts,amsmath,amssymb,amsthm}
\usepackage[bookmarks, bookmarksopen, bookmarksnumbered]{hyperref}

\usepackage{graphicx}
\usepackage{color}
\usepackage{siunitx}
\usepackage[utf8]{inputenc}
\usepackage[T1]{fontenc}
\usepackage{lmodern}
\usepackage[right]{rotating}
\usepackage{longtable}
\usepackage{array}
\usepackage{multirow}
\usepackage{paralist}
\usepackage{colortbl}

\newcommand{\codename}[1]{\texttt{#1}}

\renewcommand{\i}{\mathrm{i}}

\graphicspath{{figs/}}

\usepackage[normalem]{ulem}

\begin{document}

\title{Numerical-relativity simulations of long-lived remnants of binary neutron star mergers}
\date{\today}

\author{Roberto \surname{De Pietri}}
\affiliation{Parma University, Parco Area delle Scienze 7/A, I-43124 Parma (PR), Italy}
\affiliation{INFN gruppo collegato di Parma, Parco Area delle Scienze 7/A, I-43124 Parma (PR), Italy}

\author{Alessandra \surname{Feo}}
\affiliation{Department of Chemistry, Life Sciences and Environmental
Sustainability, Parma University, Parco Area delle Scienze, 157/A, I-43124 Parma (PR), Italy}
\affiliation{INFN gruppo collegato di Parma, Parco Area delle Scienze 7/A, I-43124 Parma (PR), Italy}

\author{Jos\'e A.~\surname{Font}}
\affiliation{Departamento de
  Astronom\'{\i}a y Astrof\'{\i}sica, Universitat de Val\`encia,
  Dr. Moliner 50, 46100, Burjassot (Val\`encia), Spain}
\affiliation{Observatori Astron\`omic, Universitat de Val\`encia, C/ Catedr\'atico 
  Jos\'e Beltr\'an 2, 46980, Paterna (Val\`encia), Spain}

\author{Frank \surname{L\"offler}}
\affiliation{Heinz Nixdorf Chair for Distributed Information Systems, Friedrich Schiller University Jena, Jena, Germany}
\affiliation{Center for Computation \& Technology, Louisiana State University, Baton Rouge, Luisiana, 70803 USA}

\author{Michele \surname{Pasquali}}
\affiliation{Parma University, Parco Area delle Scienze 7/A, I-43124 Parma (PR), Italy}
\affiliation{INFN gruppo collegato di Parma, Parco Area delle Scienze 7/A, I-43124 Parma (PR), Italy}

\author{Nikolaos \surname{Stergioulas}}
\affiliation{Department of Physics, Aristotle University of Thessaloniki, Thessaloniki 54124, Greece}

\begin{abstract}
We analyze the properties of the gravitational wave signal emitted
after the merger of a binary neutron star system when the remnant
survives for more than a \SI{80}{ms} (and up to \SI{140}{ms}). We employ four
different piecewise polytropic equations of state supplemented by an
ideal fluid thermal component. We find that the
postmerger phase can be subdivided into three phases: an early
postmerger phase (where the quadrupole mode and a few subdominant
features are active), the intermediate postmerger phase (where only
the quadrupole mode is active) and the late postmerger phase (where
convective instabilities trigger inertial modes). The inertial modes
have frequencies somewhat smaller than the quadrupole modes. In one
model, we find an interesting association of a corotation of the
quadrupole mode in parts of the star with a revival of its
amplitude. The gravitational wave emission of inertial modes in the
late postmerger phase is concentrated in a narrow frequency region
and is potentially detectable by the planned third-generation
detectors. This allows for the possibility of probing not only the
cold part of the equation of state, but also its dependence on
finite temperature. In view of these results, it will be important
to investigate the impact of various type of viscosities on the
potential excitation of inertial modes in binary neutron star merger
remnants. 
\end{abstract}

\LTcapwidth=\columnwidth


\maketitle

\section{Introduction}
\label{sec:intro}

The landmark joint detection of gravitational waves (GWs) and
electromagnetic waves from the binary neutron star merger event
GW170817 / GRB~170817A / AT~2017gfo is widely regarded as the start of
multimessenger astrophysics (see~\cite{Abbott:2017b} and references
therein). This unique event has provided key evidence to address
long-standing issues in relativistic astrophysics such as the origin
of short gamma-ray bursts and kilonovae, the $r$-process-mediated
nucleosynthesis of heavy elements, and independent measures of
cosmological parameters ~\cite{Abbott:2017e,Kasen:2017,Abbott:2017d,
  Pian:2017}. Constraints on the binary parameters, such as component
masses, radii, spins, and the equation of state (EOS), as derived from
the detected GW signal in the inspiral phase, were first presented
in~\cite{Abbott:2017a} and subsequently improved
in~\cite{Abbott:2018c,Abbott:2018d}. See
also~\cite{De:2018,Fattoyev:2018,Annala:2018,LIGOScientific:2019eut}
for additional estimates, \cite{2019arXiv190511212T} for a different
assessment of EOS constraints, and
\cite{2019PhRvD.100b3012C,2019ApJ...880L..15M} for future prospects.
The constraints on radii and EOS are based on the analysis of the
tidal interactions between the two stars during their
inspiral. Additional constraints on the radius can be obtained
assuming that there was no prompt collapse of the remnant (see
e.g. \cite{2017ApJ...850L..34B} and
\cite{2019arXiv190106969B,2019arXiv190708534B} for reviews and
references therein). Complementary information on the internal
structure of neutron stars is expected to become available through the
observation of the postmerger GW signal, primarily through the
application of an empirical relation between the $m=2$ $f$-mode
frequency and the neutron star radius (see e.g.
\cite{ShibataUryu2002,2005PhRvL..94t1101S,Stergioulas:2011gd,Bauswein:2011tp,Bauswein:2012ya,bauswein:2015unified,Clark:2015zxa,2019PhRvL.122f1102B,2019PhRvD..99d4014T,BSJ,2017arXiv171100040C,Bose:2017jvk,Yang:2017xlf}
and \cite{2019arXiv190106969B,2019arXiv190708534B} for reviews and
references therein). A first potential approach for inferring
information on the finite-temperature dependence of the EOS, based on the long-term GW emission in the postmerger
phase, was presented in \cite{DePietri:2018}.

Reliable theoretical models of postmerger GW emission can only be
obtained through nonlinear, general-relativistic numerical
simulations. Generically, such numerical simulations of binary neutron
star (BNS) mergers show that the outcome depends on the component
masses and the EOS (see
e.g.~\cite{Duez2010,FaberRasio2012,Paschalidis:2016agf,Rezzolla:2017}
for reviews). Prompt collapse to a black hole is the likely outcome
only if the initial masses are sufficiently large, while delayed
collapse (or no collapse) is produced otherwise. In
the latter cases, the resulting object may be either a hypermassive
neutron star (HMNS)~\cite{Baumgarte:1999cq}, where the remnant is
temporarily supported against collapse by
differential rotation and thermal gradients, or even a supramassive
neutron star, when the mass of the remnant is sufficiently small to be
supported by rigid rotation~\cite{CST92}. Either case leads to
sufficiently long-lived remnants to produce large amounts of
high-frequency ($\sim$kHz) gravitational
radiation. Hypermassive remnants may avoid collapse
for timescales of tens to hundreds of milliseconds. Supramassive
neutron stars will avoid collapse on an immensely longer timescale
set by magnetic dipole radiation. On such long timescales,
additional GW emission mechanisms may become relevant, such as the
Chandrasekhar-Friedman-Schutz (CFS) instability \cite{PhysRevLett.24.611,1978ApJ...222..281F}, 
shear instabilities~\cite{Watts:2002ik,Watts:2003nn,Corvino:2010} or
convective instabilities in rotating stars with non-barotropic thermal
profiles~\cite{Camelio:2019}.  There is thus good reason to study in
more detail the long-term evolution of BNS remnants.

Due to their high frequency, the GW signals produced by postmerger
remnants are much more difficult to detect for the Advanced LIGO and
Virgo detectors than those from the late inspiral phase, as the
(quantum) shot-noise limits the sensitivity of interferometers above a
few kHz. Not surprisingly, the recent searches for GWs from the
postmerger remnant conducted by the LIGO/Virgo collaboration in the
data of GW170817 have not revealed any detection ~\cite{Abbott:2017c,Abbott:2018a,Abbott:2018b}. 
Nevertheless, postmerger GWs for
GW170817-like events, in terms of distance and amplitude, seem within
reach of future observing runs with expected higher sensitivity. We
note that a seconds-long postmerger signal candidate has been
reported by~\cite{vanPutten:2019} with a lower GW energy estimate than
that computed by~\cite{Abbott:2017c}.

Simulations of the postmerger phase of HMNSs show the emission of
significant amounts of gravitational radiation at distinct frequencies
of a few kHz
(e.g.~\cite{Zhuge1994,PhysRevD.61.064001,Oechslin2002PhRvD..65j3005O,ShibataUryu2002,Shibata:2005,ShibataTani:2006,Kiuchi:2009,Stergioulas:2011gd,Bauswein:2011tp,Bauswein:2012ya,hotokezaka:2013remnant,Takami:2014tva,bauswein:2015unified,bernuzzi:2015modeling,dietrich:2015numerical,Dietrich:2016lyp,Dietrich:2016hky,Rezzolla:2016nxn,Lehner2016arXiv160300501L}),
with contributions as low as $\sim$1 kHz, depending on the
EOS~\cite{Maione:2017aux}, see \cite{Baiotti:2019sew}
  for a review and references therein. All of these simulations last
for up to a few tens of ms after merger, which is long enough to show
that the GW emission is dominated by a main peak in
the frequency range $\sim$2-4 kHz, with secondary peaks
on both sides of the main peak.
Employing a mode-analysis technique it was found in~\cite{Stergioulas:2011gd}, that the main postmerger peak is due to the excitation of the fundamental $m=2$ $f$-mode
(denoted in the literature as or $f_{\rm peak}$ or $f_2$). In
addition, it was shown in \cite{Stergioulas:2011gd} that
some of the secondary peaks can be explained as a
quasi-linear combinations between $f_2$ and the fundamental,
quasi-radial, $m=0$ mode (that is, sums and differences of these
frequencies, denoted e.g. as $f_{2-0}$ and $f_{2+0}$). These
combination frequencies are primarily present in models that are
relatively close to the threshold mass to collapse (high-mass/soft EOS
models) \cite{bauswein:2015unified}. In addition,
in~\cite{bauswein:2015unified} it was found that a different secondary
peak, $f_{\rm spiral}$, is present in models that are relatively far
from the threshold mass to collapse (low-mass/stiff EOS) and it is due
to a spiral deformation excited during merger. Depending on the
relative strength between $f_{2-0}$ and $f_{\rm spiral}$,
\cite{bauswein:2015unified} introduced a spectral classification of
the postmerger GW emission (see \cite{BSJ,2019arXiv190106969B} for
reviews).
For additional studies of the properties of postmerger remnants (where sometimes a different 
label for the mode is used), see
e.g.~\cite{Takami:2014tva,Rezzolla:2016nxn,2008PhRvD..78h4033B,2008PhRvD..77b4006A,2008PhRvD..78b4012L,Kastaun:2014fna,DePietri:2015lya,Kastaun:2016yaf,PEPS2015,EPP2016,Hanauske:2016gia,Foucart:2015gaa,Maione:2017aux,2017PhRvD..95l3003S,2018ApJ...860...64F}
as well as \cite{2019arXiv190708534B} and references therein.

Recently we performed numerical simulations in full general relativity
of BNS systems, extending the simulation time significantly longer
than those previously reported in the literature~\cite{DePietri:2018}.
These simulations allowed us to find new features
in the GW spectrum of the remnant on longer
timescales.  In particular, the simulations were extended up to
$\sim$140 ms after merger and were based on a piecewise polytropic
approximation for the EOS treatment, supplemented by
a thermal component. We found that after an initial phase of a few
tens of ms (where the dominant oscillation mode of the
remnant and main GW emitter is the $m=2$ $f$-mode),
convective instabilities in the remnant may excite
inertial modes, which also radiate GWs. The phase during which
inertial modes of global and discrete nature can be excited, may
last at least up to several tens of milliseconds
and, as shown in~\cite{DePietri:2018}, the associated
 GW emission may become potentially observable by the planned
third-generation GW detectors at frequencies of a few kHz.

Here, we complement the study initiated in~\cite{DePietri:2018} by
carrying out the analysis of additional simulations, which account for
two new EOS, namely H4 and MS1.  Within the range of validity of the
input physics and assumptions employed in our simulations, the
extended results presented here confirm our earlier findings in a
generic way. 
we present a convergence check for a specific model to further
validate our results.

The paper is organized as follows: in Section~\ref{sec:setup} we
discuss our initial models and numerical
setup. Section~\ref{sec:results} presents our main results, with a
particular attention to the mode analysis in
Sections~\ref{subsec:Spectrograms},~\ref{subsec:Spectra},
~\ref{subsec:wavefunction} and~\ref{subsec:instability}. A summary of the
main results presented in the previous sections of this work is
provided in Section~\ref{sec:conclusions}. The paper closes with three
appendices, where numerical issues and convergence properties of our
simulations are discussed (Appendix A), along with a revision of past literature regarding 
long-term simulations of the postmerger phase of binary neutron star mergers (Appendix B), 
and the thermodynamical properties of the piecewise-polytropic EOS used in this study 
(Appendix C).

Unless otherwise noted, we employ units of $c=G=M_\odot=1$.

\begin{figure}
\includegraphics[width=0.45\textwidth]{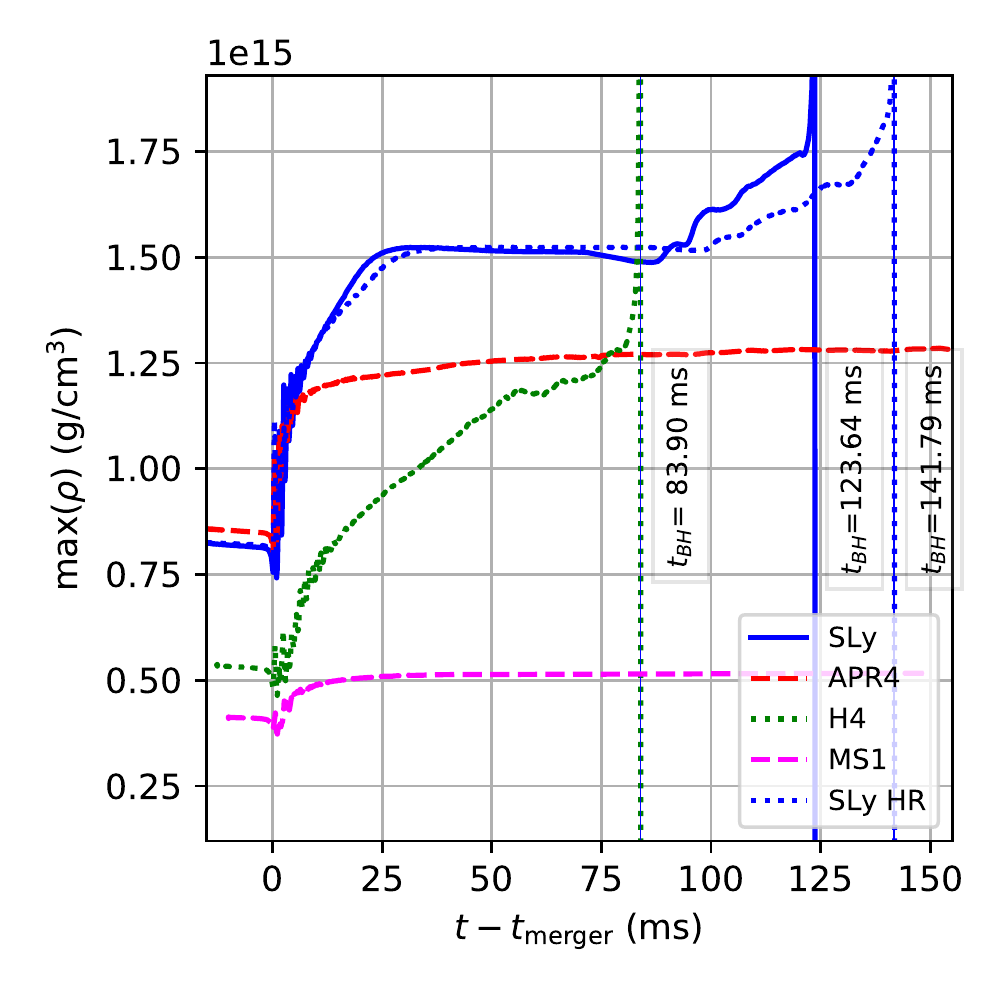}
\caption{Maximum density as a function of time
  for the simulations of the models with the four different EOS (notice that HR is a high-resolution run with the SLy EOS).}
\label{fig:maxdens}
\end{figure}

\begin{figure*}
\includegraphics[width=0.95\textwidth]{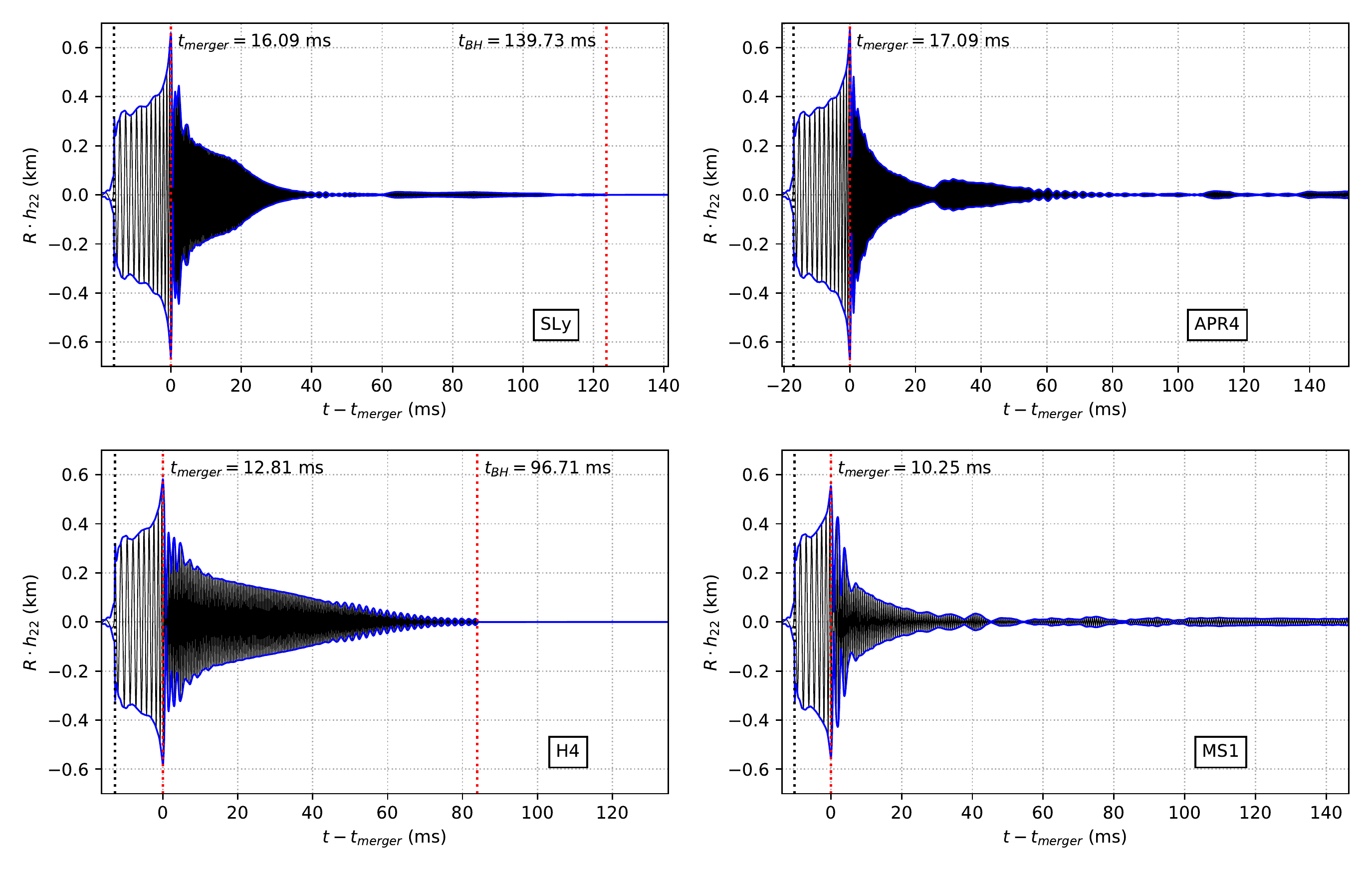}
\vspace{-6mm}
\caption{ Scaled $h_{22}$ component of the GW waveform for the  four models. 
The red dotted lines marks the merger time, $t_{\rm merger}$, and the time of the formation
of the black hole, $t_{BH}$.}
\label{fig:Waveforms}
\end{figure*}

\section{Initial data and setup}
\label{sec:setup}

\begin{figure*}
\includegraphics[width=0.95\textwidth]{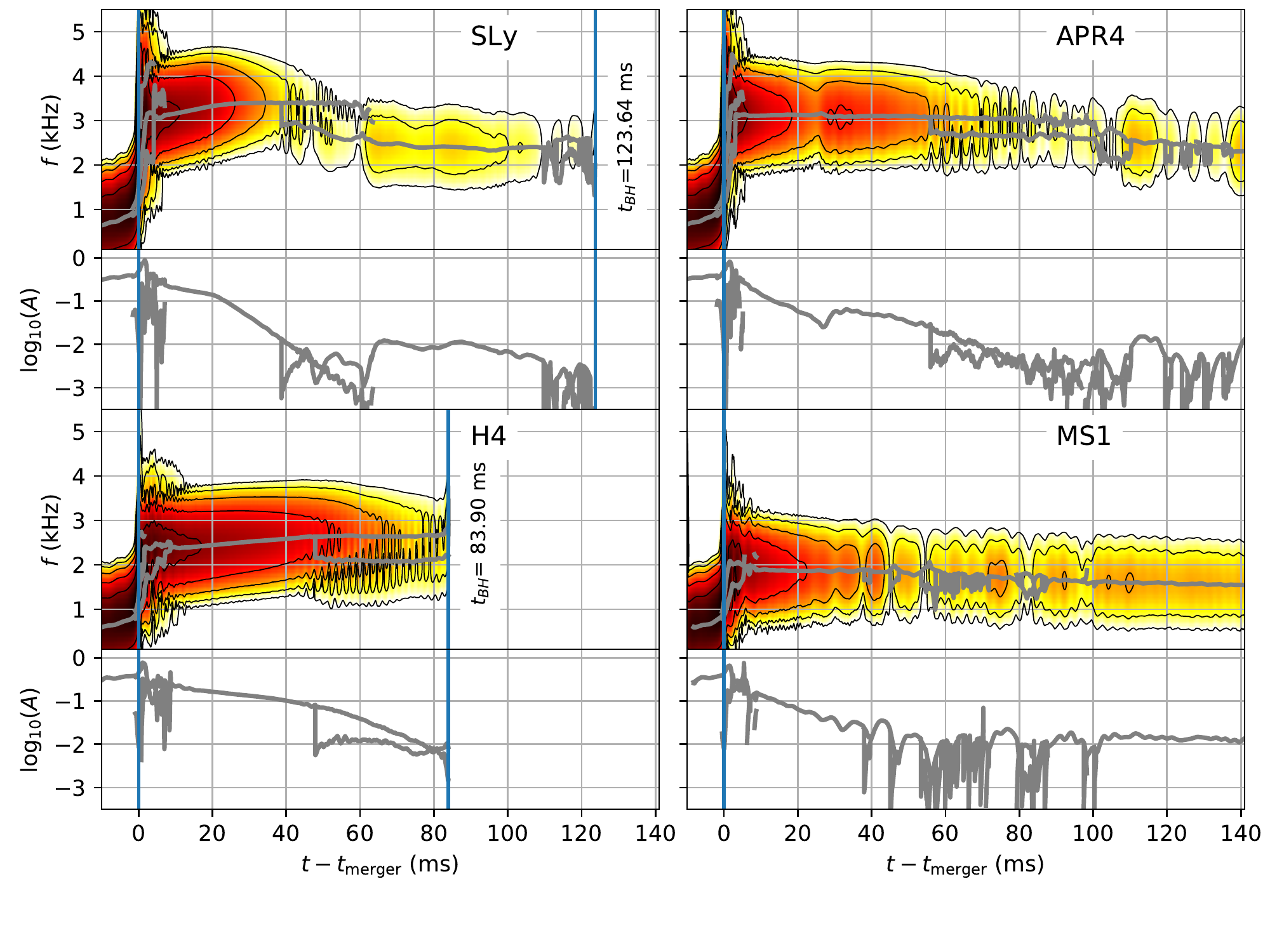}
\vspace{-13mm}
\caption{The top half of each panel shows time-frequency spectrograms
of the $l=m=2$ component of the GW strain for all models.  Thick
gray lines indicate the frequency of the active
modes that mainly responsible for the GW emission at different
times. Colors indicate the relative intensity of the spectral
density (darker areas correspond to higher intensity). The bottom
half of each panel displays the amplitude of the main active modes
(in arbitrary units). For the two models that collapse to a black
hole within the simulated timescale, the time of black hole
formation is reported with respect to the merger time and indicated
by the rightmost vertical lines.}
\label{fig:MAIN}
\end{figure*}

Initial data for irrotational neutron star binaries in the conformal
thin sandwich approximation are generated using the \codename{LORENE}
code~\cite{Lorene:web,Gourgoulhon:2000nn}. We employ four EOS, namely
SLy, APR4, H4, and MS1, parametrized as piecewise
polytropes~\cite{Read:2009constraints} with 7 pieces plus a thermal
component with adiabatic index $\Gamma_{\rm th}=1.8$. The main
properties of our initial data (see Table I) are a total baryonic mass of
2.8~M$_{\odot}$ and a total mass of $\simeq 2.55-2.60~M_\odot$ with an
initial separation of $\approx$ \SI{44.3}{km} (four full orbits before
merger). We consider equal-mass systems characterized by relatively
low-mass components, below the range of the inferred masses for
GW170817~\cite{Abbott:2018c} (which had a total mass between 2.73
M$_{\odot}$ for low-spin priors and 2.77 M$_{\odot}$ for high-spin
priors).

\begin{table}
\begin{center}
\begin{tabular}{r@{\hskip 1.5em}ccc@{\hskip 1.5em}cc@{\hskip 1.5em}c}
EOS  & $M_0$ & $M$  &  $C$  & $M_\mathrm{ADM}$ & $J_\mathrm{ADM}$ & $\Omega_0$ (krad/s)\\
\hline
SLy   &  1.40  & 1.2810 & 0.161 & 2.538 & 6.623 & 1.770   \\
APR4  &  1.40  & 1.2755 & 0.166 & 2.528 & 6.577 & 1.767   \\
H4    &  1.40  & 1.3004 & 0.137 & 2.576 & 6.802 & 1.783   \\
MS1   &  1.40  & 1.3047 & 0.129 & 2.585 & 6.850 & 1.787   \\
\end{tabular}
\vspace{-2mm}
\end{center}
\caption{Main properties of the four equal-mass BNS systems studied in
  this work. The columns report the baryonic mass, $M_0$, the
  gravitational mass, $M$, and the compactness $C:=M/R$ of the
  individual stars, the total ADM mass, $M_\mathrm{ADM}$, the angular
  momentum, $J_\mathrm{ADM}$, and the angular velocity of the binary
  system, $\Omega_0$, at the start of the simulation. All systems have
  roughly the same initial separation of 44.3 km.}
\label{tab:ID}
\end{table}

The initial data are evolved using the \codename{Einstein
  Toolkit}~\cite{Loffler:2011ay}, an open source, modular code for
numerical relativity based on the Cactus
framework~\cite{Cactuscode:web,Goodale:2002a}. For the simulations we
employ the same setting in the code as
in~\cite{DePietri:2015lya,Maione:2016zqz,Feo:2016cbs,Maione:2017aux,2019ApJ...881..122D}. The
only difference here is the use of $\pi$-symmetry to reduce the
computational cost by a factor 2. This allows to push the limit of the
simulated time to $\sim$\SI{150}{ms}, of which the last \SI{130}{ms}
correspond to the postmerger phase.

The set of equations we solve numerically comprise Einstein's
gravitational field equations, the general relativistic hydrodynamics
equations for a perfect-fluid stress-energy tensor
and the relativistic energy conservation
equation. Therefore, our simulations do not account for effects due
to magnetic fields or non-thermal radiation transport associated with
neutrinos. In particular, we evolve Einstein's equations in the BSSN
formulation~\cite{Shibata:1995we,Baumgarte:1998te}, as implemented in
the \codename{McLachlan} module \cite{McLachlan:web}, and the Valencia
formulation of the general relativistic hydrodynamics
equations~\cite{Banyuls:1997,Font:2008}, as implemented in the public
\codename{GRHydro} module~\cite{Baiotti:2004wn,Moesta:2013dna}. For
the matter fields we employ a finite-volume algorithm with the HLLE
Riemann solver~\cite{Harten:1983on,Einfeldt:1988og} and the WENO
reconstruction method \cite{liu1994weighted,jiang1996efficient}. The
combined use of WENO reconstruction and the BSSN formulation was found
in~\cite{DePietri:2015lya} to be the best combination within the
\codename{Einstein Toolkit} even at low resolution. For the time
evolution we used the Method of Lines with a fourth-order,
conservative Runge-Kutta scheme~\cite{Runge:1895aa,Kutta:1901aa}. The
evolved variables for both the spacetime and the hydrodynamics are
discretized on a Cartesian grid with 6 levels of fixed mesh
refinement, each using twice the resolution of its parent level. The
outermost boundary of the grid is set at $720 M_{\odot}$ ($\approx
\SI{1040}{km}$) from the center. The standard spatial resolution at the
finest refinement level in our production runs is $dx=\SI{277}{m}$.  For the
SLy EOS, we also use a higher-resolution setup, with spatial
resolution at the finest level of $dx=\SI{185}{m}$ (this simulation is
denoted as SLy HR below). A comparison of GW\ spectrograms for the two
resolutions is provided in Appendix A.
 
\section{Results}
\label{sec:results}

\subsection{Matter dynamics and waveforms}
\label{sec:dynamics}

The dynamics of the matter of the four models considered in this work
can be followed by analyzing the time evolution of
quantities such as the central density and the density
distribution within the equatorial and the vertical planes. The overall
behavior for the first phase (inspiral and merger) of the binary
system's evolution is similar for each model, until
the two neutron stars form a bar-deformed remnant as in
\cite{DePietri:2015lya} (the appearance of a bar
deformation is due to the rotating pattern of the dominant $m=2$
$f$-mode excited after merger). During the initial
postmerger phase we note some oscillations in the
evolution of the maximum density, as shown in
Figure~\ref{fig:maxdens}, which have different durations for each
model. For the SLy and MS1 EOS these oscillations last $\sim$7 ms
while for the APR4 and H4 $\sim$4 and $\sim$5 ms respectively, until
the core becomes more stable and the bar-deformed remnant is
left. This configuration of a bar-deformed remnant survives for some
tens of milliseconds for each model before becoming
nearly axisymmetric and more stable.

We note that for the SLy and H4 EOS the density of the core grows even
after the oscillations in the maximum density have
dampened (see Sec. \ref{subsec:Spectrograms} for a discussion of the
frequency evolution of the dominant $m=2$ $f$-mode during this time
interval). Eventually, a delayed collapse to a black hole occurs for
both of these two models, at $\sim$124 and $\sim$84 ms after the
merger, respectively, while we do not observe a collapse to a black
hole in the case of APR4 and MS1, before we terminate
  our simulations at $t-t_{\rm merger}\sim \SI{150}{ms}$.

The GW signal obtained from the simulations of these four models is
presented in Figure~\ref{fig:Waveforms}. We show the
$h_{22}$ component of the GW amplitude (multiplied by the assumed
distance to the source, $R$) (black line) and the absolute value of
$h_{22}$ (blue line), assuming an ideal source orientation.  The
time of merging and the time of the collapse to a black hole are
indicated by vertical, dotted red lines. For
details on the numerical extraction of the GW signal see
~\cite{DePietri:2015lya,Maione:2016zqz,Maione:2017aux}.

The evolution of the GW amplitude for the SLy EOS
model in Fig.~~\ref{fig:Waveforms} shows a relatively slow decay
up to about \SI{20}{ms} after merger (in this phase, the decay is due to
both nonlinear hydrodynamical interactions and GW emission),
followed by a second phase of somewhat faster decay, which lasts up
to about 65 ms after merger, when the GW emission almost ceases and
the remnant has become nearly axisymmetric.  However, at about \SI{65}{ms}
after merger, there is a clear revival of GW emission (see
~\cite{DePietri:2018} and the following sections, for an
interpretation of this observation in terms of the excitation of
inertial modes by convective instabilities in the remnant).

For the model with\ EOS MS1, we observe an initial
decay of GW emission up to about \SI{40}{ms} after merger, followed by
several episodes of revival of GW emission, up to about \SI{100}{ms} after
merger. At even later times, the remnant radiates GWs at
practically a single frequency and at a practically constant
amplitude. Such a behavior is expected in the case of a specific,
unstable mode saturating at a maximum amplitude, due to a balance
between energy gained (by an instability) and energy lost (e.g. by
GW emission).  

For the model with EOS H4, Fig.~~\ref{fig:Waveforms} reveals an 
almost linear decay of the GW amplitude for the most part of the 
time period up to almost \SI{80}{ms} (shortly after, the remnant 
collapses to a black hole).  However, after \SI{45}{ms} from merger, 
a modulation appears in the amplitude evolution, indicating the 
presence of a second frequency, close to the dominant postmerger 
frequency (in ~\cite{DePietri:2018} and the following sections, 
we give the same interpretation to this second
frequency, as in the case of the late-time behavior of the SLy EOS
model).

A similar late-time revival (more than \SI{60}{ms} after
merger) of GW emission can be observed for the model with the APR4
EOS. In addition, for this model we observe a strong revival of GW
emission already at 25~ms after merger. We will comment on this
finding in Section \ref{subsec:corot}.

\subsection{Spectrum evolution and Prony's analysis}\label{subsec:Spectrograms}

The general dynamics of the evolution of our four BNS systems can be
neatly captured in the time-frequency plots of the $h_{22}$ component
of the spherical harmonic decomposition of the GW signal shown in
Figure~\ref{fig:MAIN}. On top of each spectrogram we also superimpose
with thick gray lines the time evolution of the frequency of the main
active spectral {modes} of the remnant. This is determined using the
ESPRIT Prony's method (employing a moving window interval of
\SI{3}{ms}) as discussed in~\cite{Maione:2017aux}. The corresponding
extracted amplitudes of the modes are also reported in the lower half
of each panel in arbitrary units. The leftmost vertical line in each
panel of Figure~\ref{fig:MAIN} indicates the time of merger and the
rightmost vertical line (for EOS SLy and H4 only) indicates the time
of black hole formation.

For the H4 EOS and the SLy EOS, the postmerger remnant survives for
over 80 ms and 120 ms, respectively, before collapsing to a black
hole. For the MS1 EOS and APR4 EOS, the remnant has not collapsed even
after 140 ms. The findings about the mode dynamics presented
in~\cite{DePietri:2018} continue being valid when adding the H4 and
MS1 EOS to the sample. For all four EOS the dominant mode in the early
postmerger phase is the $m=2$ $f$-mode, with a frequency above 3 kHz
for SLy and APR4, about 2.5 kHz for H4, and below 2 kHz for MS1. In
all four cases, the evolution of this mode is characterized by a
decaying amplitude. As first found by~\cite{DePietri:2018}, distinct,
lower-frequency modes appear later in the evolution, at $t-t_{\rm
  merger}\sim$35 ms for SLy and MS1, $\sim$45 ms for H4, and $\sim$55
ms for APR4. This new type of GW emission dominates over the initial
$m=2$ $f$-mode at late times.

A closer look at our Prony's analysis of the behavior of the
postmerger signal for each EOS reveals the following (all times are
w.r.t. the time of merger):
 
For the \underline{SLy EOS} the early postmerger phase (which
includes oscillations in the maximum density and transient GW emission
of secondary peaks) ends at about \SI{5}{ms}. After this, the $m=2$
$f$-mode is practically the only significant oscillation mode and we
observe that its frequency changes secularly with time, at a constant
rate of $\sim$\SI{1.6e-2}{kHz/ms} up to $\sim$\SI{20}{ms}. From then
on and up to about \SI{60}{ms} the frequency of the $m=2$ $f$-mode remains
practically constant. The decay of the amplitude of the $m=2$ $f$-mode
is slower in the first \SI{20}{ms} than in the period between \SI{20}{ms} and \SI{60}{ms}
(at the end of this period this mode has practically faded away). At
$\sim$\SI{37}{ms} an additional mode at a different (lower) frequency
appears alongside with the $m=2$ $f$-mode. The frequency of the new
excited mode is about $\sim$\SI{2.93}{kHz}. Within about \SI{8}{ms} from its
appearance, the new mode has grown more than an order of magnitude in
amplitude and then its amplitude rapidly collapses. Another phase of
the excitation, saturation and destruction of a distinct,
low-frequency mode (with frequency of roughly \SI{2.7}{kHz}) follows in the
time period between \SI{45}{ms} and \SI{60}{ms}. A third such phase lasts between
\SI{60}{ms} and \SI{110}{ms}, where the dominant mode has a nearly constant
frequency of about \SI{2.5}{kHz} and a nearly constant amplitude (with small
variations). Notice that the saturation amplitude of the low-frequency
modes in these three periods is similar in order of magnitude. In the
following sections (as in ~\cite{DePietri:2018}) we interpret the
low-frequency modes as inertial modes excited by convective
instabilities in the remnant (in the model with the SLy EOS we thus
observe three such episodes of inertial mode excitation, with somewhat
different frequencies, but comparable saturation amplitude - whether
it is the same or different inertial modes that are excited in each
episode remains to be determined by a detailed mode analysis).  At
$\sim$\SI{110}{ms} it is not possible to recognize any mode clearly
and at $\sim$\SI{123.6}{ms} a black hole is formed.
 
For the \underline{APR4 EOS} the early postmerger phase ends at about
~\SI{3.2}{ms}. At this point there is only one mode with frequency
$\sim$\SI{3.1}{kHz} which remains practically constant up to
$\sim$\SI{25}{ms}.  The exponential damping timescale is constant in
this time interval. At $\sim$\SI{25}{ms} there is a small glitch in
the frequency of the main mode, which otherwise continues practically
constant up to $\sim$\SI{54}{ms}. Notice that at $\sim$\SI{25}{ms} the
amplitude of the mode increases appreciably on a dynamical timescale,
before the damping continues (we will comment on this finding in
Section \ref{subsec:corot}).  The frequency of the main mode continues
to be visible up to $\sim$\SI{100}{ms}, having a somewhat larger
damping rate.  However, at $\sim$\SI{54}{ms} an additional mode
appears, with a lower frequency of $\sim$\SI{2.70}{kHz}. Whereas its
frequency remains practically constant up to $\sim$\SI{100}{ms}, its
amplitude shows rapid variations, which can be interpreted as several
episodes of excitation-saturation-destruction of an unstable
mode. Between $\sim$\SI{105}{ms} and $\sim$\SI{120}{ms} another such
episode takes place (at the same frequency), but this time the
saturation amplitude rises higher and remains constant for a longer
time. After $\sim$\SI{120}{ms} a few more episodes are visible at a
frequency slightly smaller than in the previous time period. Notice
that between $\sim$\SI{98}{ms} and $\sim$\SI{105}{ms} another mode
with smaller frequency of $\sim$\SI{2}{kHz} appears in the spectrogram,
but with very small amplitude (we cannot exclude that this could be an
artifact of the Prony's method).

For the \underline{H4 EOS} the early postmerger phase ends at
$\sim$\SI{6.7}{ms} when only one mode is present. This mode has an
initial frequency of $\sim$\SI{2.37}{kHz} and changes constantly with
a rate of $\sim$\SI{7.1e-3}{kHz/ms} up to $\sim$\SI{46}{ms} after the
merger, after which it remains practically constant at
$\sim$\SI{2.64}{kHz}. The exponential damping timescale is constant in
this time interval (and becomes somewhat shorter after that). At
$\sim$\SI{46}{ms} a second mode appears at a practically constant,
lower frequency of $\sim$\SI{2.11}{kHz}.  This new modes rises in
amplitude on a dynamical timescale, saturates quickly and stays at a
nearly constant amplitude until black hole formation at
$\sim$\SI{84}{ms}.
 
For the \underline{MS1 EOS} the early postmerger phase ends at
$\sim$\SI{6.7}{ms}, when only one single mode is active, with an
initial frequency of $\sim$\SI{1.88}{kHz} that stays practically
constant up to $\sim$\SI{35}{ms}. The exponential damping timescale is
constant in this time interval. From $\sim$\SI{35}{ms} to
$\sim$\SI{100}{ms} one can observe several episodes of the
excitation-saturation-destruction of another, lower-frequency mode. As
time goes on, the frequency of this second mode decreases. The final
episode takes place at $\sim$\SI{100}{ms}, after which the
lower-frequency mode maintains a nearly constant (but slightly
decreasing) frequency of $\sim$\SI{1.55}{kHz} and a nearly constant
saturation amplitude.

To summarize, for all four models the postmerger phase can be
subdivided into three distinct periods:
\begin{itemize}
\item {\it Early postmerger phase} ($0< t-t_{\rm merger}\lesssim$ 3-\SI{7}{ms}): the main $m=2$ $f$-mode frequency and the subdominant frequencies $f_{2-0}$ and/or $f_{\rm spiral}$ are active.
\item {\it Intermediate postmerger phase} ( 3-\SI{7}{ms} $\lesssim t-t_{\rm merger}\lesssim$ 35-\SI{50}{ms}): only the $m=2$ $f$-mode is active.
\item {\it Late postmerger phase} ($t-t_{\rm merger}\gtrsim$ 35-\SI{50}{ms}): low-frequency modes appear in several episodes of excitation-saturation-destruction.
\end{itemize}
In addition, we notice a sudden revival of the $m=2$ $f$-mode  in the intermediate phase (see discussion in Section \ref{subsec:corot}).

In the following sections, as in \cite{DePietri:2018}, we will
identify the low-frequency modes in the late postmerger phase as
inertial modes, driven by convective instabilities.

\begin{figure*}
\includegraphics[width=0.95\textwidth]{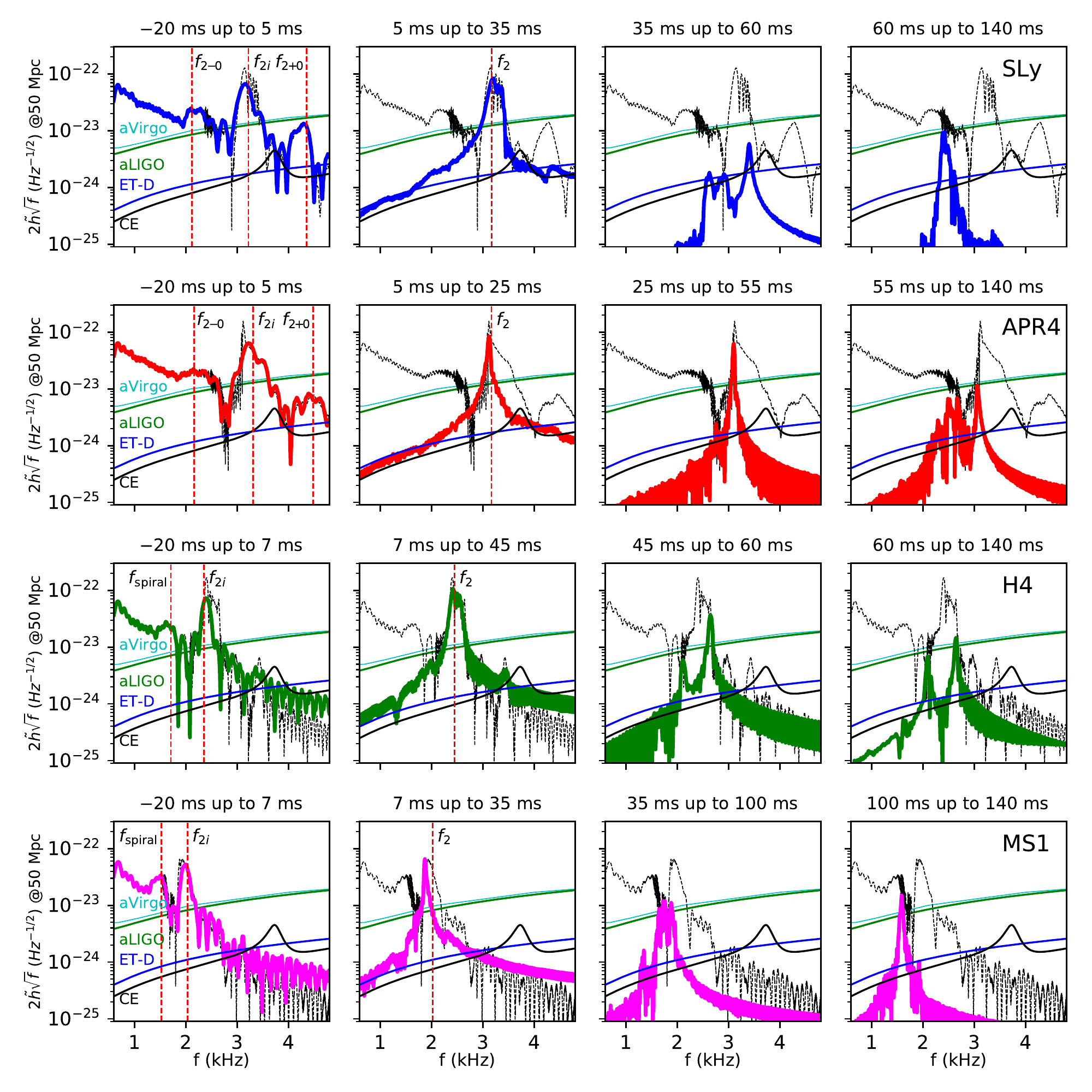}
\caption{GW spectra for a BNS merger at 50 Mpc with optimal orientation {(one row for each EOS model)}.  The spectra are shown for the entire GW signal (thin-dashed lines) and for different restricted time windows (indicated at the top of each frame) in order 
to  {emphasize the contribution }of the dominant spectral components {at different times.} The design sensitivities of Advanced Virgo~\cite{TheVirgo:2014hva}, 
Advanced LIGO~\cite{TheLIGOScientific:2014jea}, Einstein Telescope~\cite{Punturo:2010zz}, and Cosmic Explorer~\cite{Evans:2016mbw} 
are shown for reference.}
\label{fig:SPECTRUM}
\end{figure*}

\subsection{Spectral content in the different stage of the dynamics}\label{subsec:Spectra}

According to the temporal subdivisions discussed in the previous
section, it is meaningful to single out the particular contributions
of different time windows in the GW spectra for each of the EOS. The
corresponding spectra are displayed in Figure~\ref{fig:SPECTRUM}. This
figure shows the {scaled PSD $2 \tilde h \sqrt{f}$ }of the GW signal
for a maximally-aligned source at a distance of 50 Mpc (where $\tilde
h$ is the Fourier transform of the GW signal). From top to bottom each
row corresponds to the spectra of the models with the SLy, APR4, H4,
and MS1 EOS, respectively. The thin-dashed-black lines in all of the
panels display the entire spectra computed taking into account the
complete temporal evolution of the simulations, i.e.~$t-t_{\rm
  merger}\in[-20,140]$ ms. On the other hand, the colored thick-solid
lines indicate the corresponding spectra for the particular time
window reported in the top of each frame. To obtain a cleaner separation
of the contributions to the spectra due to the various time intervals
considered, we do not apply any windowing functions.  As a result, the
FFT {includes small artifacts} due to the finite size of the time
intervals. As the time window changes from left to right, the more
prominent features in the spectra can be seen to readjust accordingly.

The vertical red dashed lines in the first column (roughly
corresponding to the early postmerger phase) of
Figure~\ref{fig:SPECTRUM} mark the position of the initial frequency
$f_{2i}$ of the dominant $m=2$ $f$-mode and of the subdominant peaks
$f_{2-0}, f_{2+0}$ (for the models with EOS SLy and APR4) and $f_{\rm
  spiral}$ (for the models with EOS H4 and MS1).  These peaks can be
obtained directly from the GW spectrum, but also using the ESPRIT
Prony algorithm. Our identification of the subdominant peaks is in
agreement with the spectral classification of the postmerger GW
emission in \cite{bauswein:2015unified}, see also
\cite{2019arXiv190106969B}. Notice that in other works, the $f_{2-0}$
and $f_{\rm spiral}$ peaks have been named, collectively, $f_1$ and
the $f_{2+0}$ peak has been named $f_3$.

In the second column (roughly corresponding to the intermediate
postmerger phase) of Figure~\ref{fig:SPECTRUM}, the vertical red
dashed lines mark the position of the $f_2$ frequency peaks of the
dominant $m=2$ $f$-mode (notice that $f_2$ can remain very close to
its initial value $f_{2i}$, or it can display a secular time
variation).

For all models, the spectra corresponding to the entire duration of
the simulations is clearly dominated by the $m=2$ $f$-mode ($f_2$) in
the EOS-dependent frequency range $\sim$2-3.5 kHz. On the other hand,
when selecting the signal in the different time periods displayed in
Figure~\ref{fig:SPECTRUM} the dominant peaks in the spectra begin to
change. In the final time interval, $t-t_{\rm merger}\in[60,140]$ ms,
the spectrum is dominated by a peak at frequency $\sim$2.38 kHz in the
case of the SLy EOS and well below 2 kHz for the MS1 EOS. Moreover, in
the case of the APR4 EOS there are several nearby lower-frequency
peaks contributing to the GW spectrum in the late postmerger
phase. 

In the intermediate postmerger phase, the spectra for the EOS SLy
model (\SI{5}{ms} to \SI{35}{ms}) and the EOS H4 model (\SI{7}{ms} to
\SI{45}{ms}) are consistent with the Prony's method results that, for
these models in these time periods, the dominant $f_2$ frequency is
subject to a slow drift to higher frequencies. This is due to the fact
that in this time period the remnant experiences an almost steady
drift to higher maximum densities (see Fig.~\ref{fig:maxdens}).  It
is well known that the frequency of the fundamental fluid modes of a
star scale as the square root of the average density (see
e.g. \cite{1999LRR.....2....2K}).  Notice that the remnants for these
two models have a mass relatively close to the threshold mass for
collapse to a black hole (and indeed they do collapse within the
duration of our simulations).  The panels in Fig.~~\ref{fig:SPECTRUM}
also show the design sensitivities of Advanced
Virgo~\cite{TheVirgo:2014hva} and Advanced
LIGO~\cite{TheLIGOScientific:2014jea} as well as those of
third-generation detectors Einstein Telescope~\cite{Punturo:2010zz}
and Cosmic Explorer~\cite{Evans:2016mbw}. In~\cite{DePietri:2018} we
discussed the prospects of detectability of the low-frequency modes
excited at sufficiently long times after merger for models SLy and
APR4. In the current work our simulations account for additional
models with the H4 and MS1 EOS. In all cases we find that there is
sufficient power in the lower-frequency modes to render them
potentially observable by third-generation detectors. For a source at
50 Mpc, the expected Cosmic Explorer~\cite{Evans:2016mbw} S/N ratios for
optimal use of matched filtering techniques for the signal emitted in
the time intervals used in the last two columns of
Fig.~~\ref{fig:SPECTRUM} are $\sim$1.3 and $\sim$3.5 for SLy, $\sim$9.1
and $\sim$3.7 for APR4, $\sim$11.6 and $\sim$5.1 for H4, $\sim$9.7 and
$\sim$8.3 for MS1 EOS, respectively (for the Einstein
Telescope-D~\cite{Punturo:2010zz} they are $\sim$1.2 and $\sim$2.5,
$\sim$8.0 and $\sim$2.9, $\sim$8.7 and $\sim$3.7, $\sim$6.4 and
$\sim$5.5, but enhancement can be expected due to the triangular
arrangement with three non-aligned interferometers). We note that the
S/N ratio is significantly higher in the first $\sim\SI{40}{ms}$ after merger
for each model, while it becomes almost two orders of magnitude
smaller when the inertial modes appear and the signal can only be
observed by the next generation detectors, as mentioned above.

\begin{figure}
\includegraphics[width=0.5\textwidth]{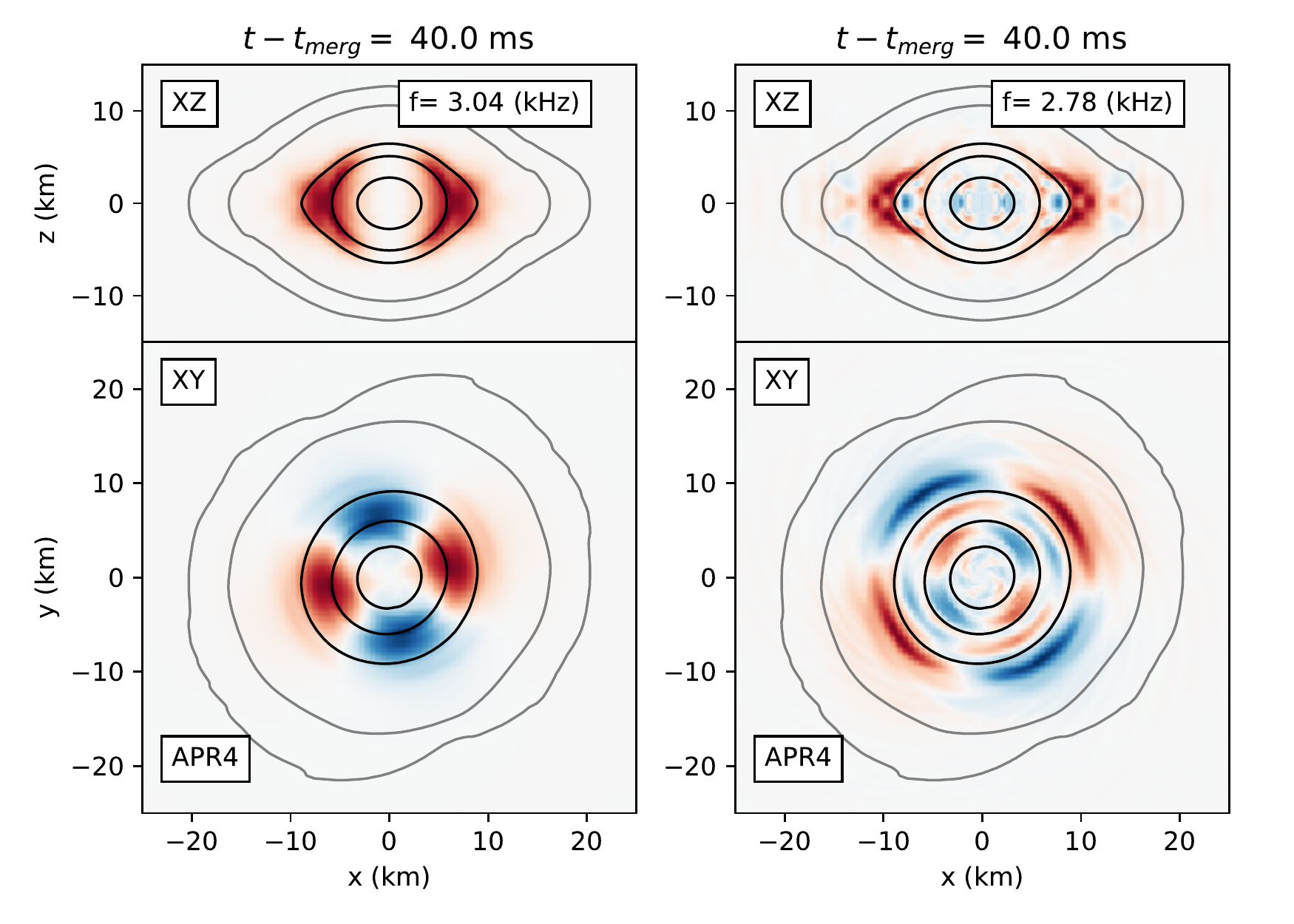}
\caption{Density eigenfunctions in the vertical and equatorial plane of the
APR4 model at 40 ms after merger. The black and gray lines are isocontours of the rest-mass
density. The left column displays the $m=2$ $f$-mode and the 
right column corresponds to an inertial mode in the late postmerger phase.}
\label{fig:APR4_mode_1}
\end{figure}

\begin{figure}
\includegraphics[width=0.5\textwidth]{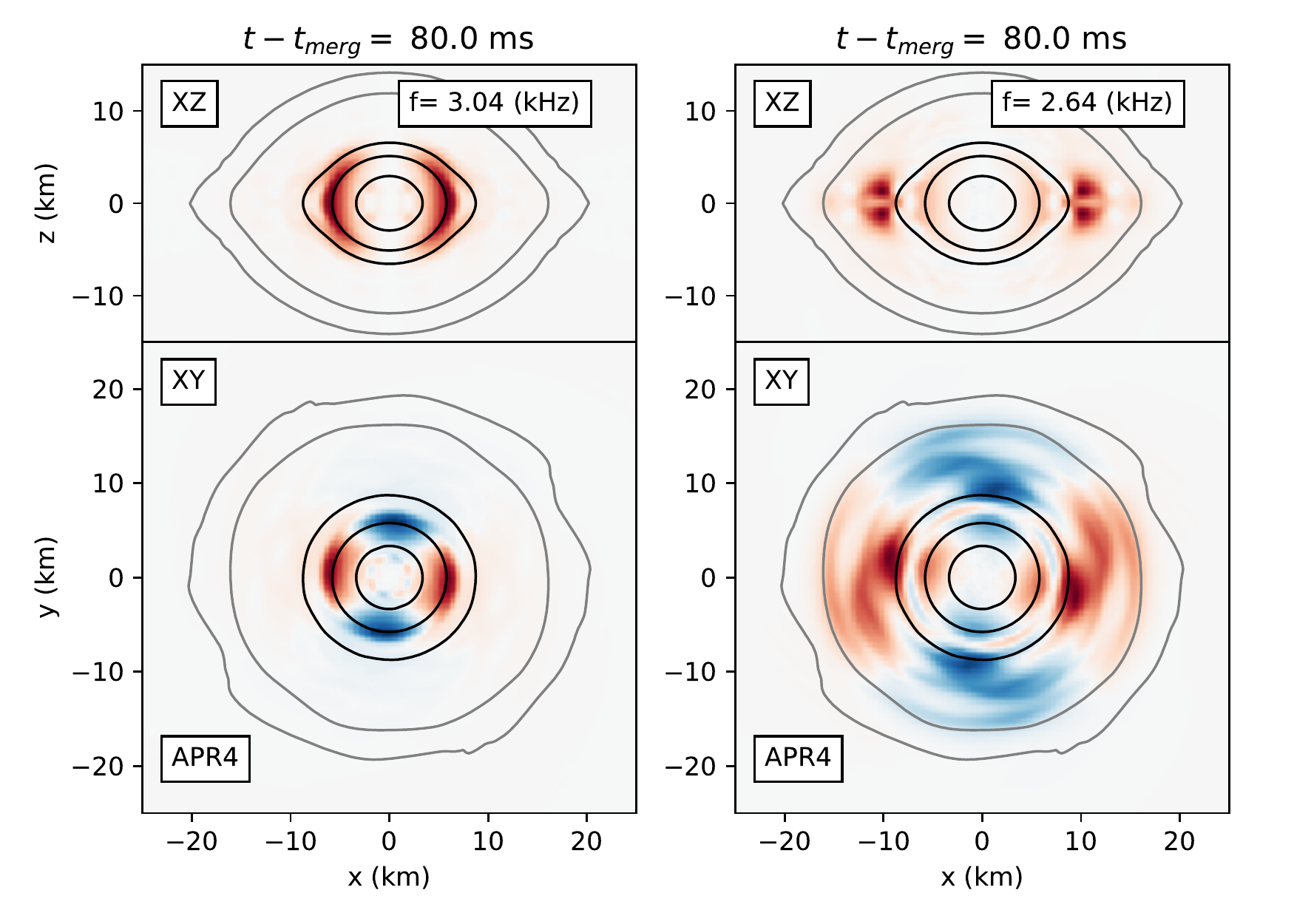}
\caption{As in Fig.~\ref{fig:APR4_mode_1}, but at \SI{80}{ms} after merger.}
\label{fig:APR4_mode_2}
\end{figure}

\begin{figure}
\includegraphics[width=0.5\textwidth]{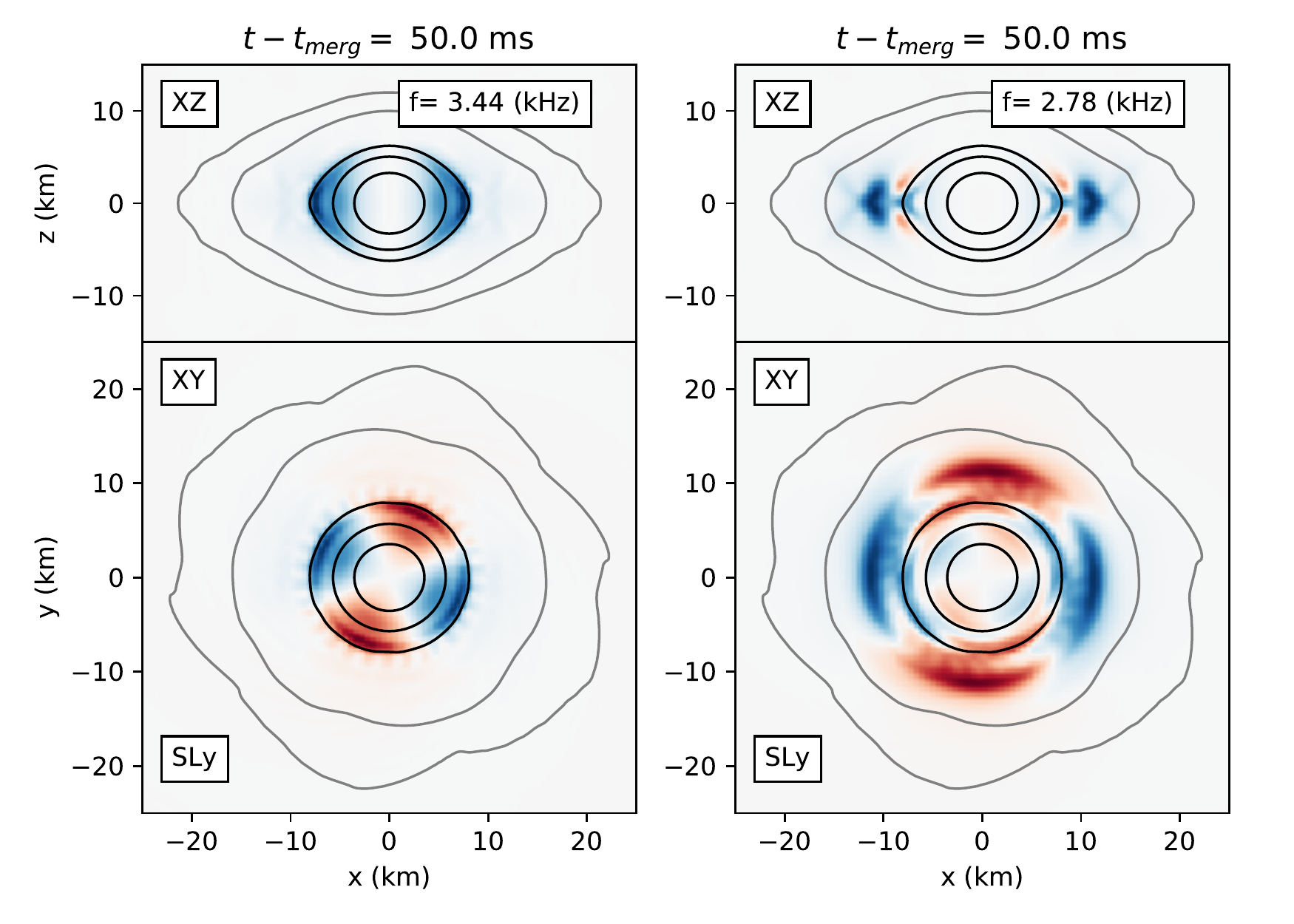}
\caption{As in Fig.~\ref{fig:APR4_mode_1}, but for EOS SLy at \SI{50}{ms} after merger. }
\label{fig:SLy_mode}
\end{figure}

\begin{figure}
\includegraphics[width=0.5\textwidth]{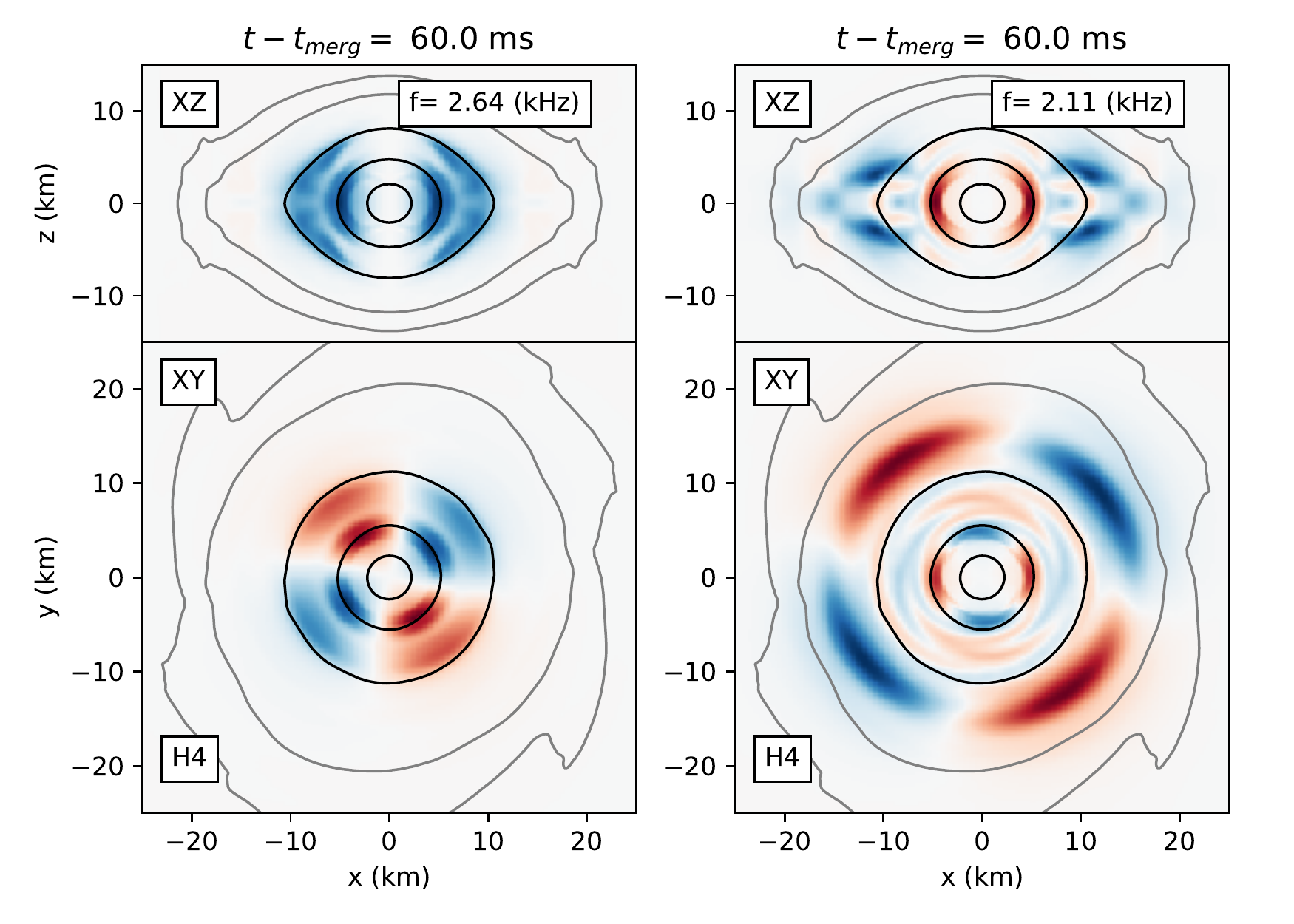}
\caption{As in Fig.~\ref{fig:APR4_mode_1}, but for EOS H4 at \SI{60}{ms} after
merger.}
\label{fig:H4_mode}
\end{figure}

\begin{figure}
\includegraphics[width=0.5\textwidth]{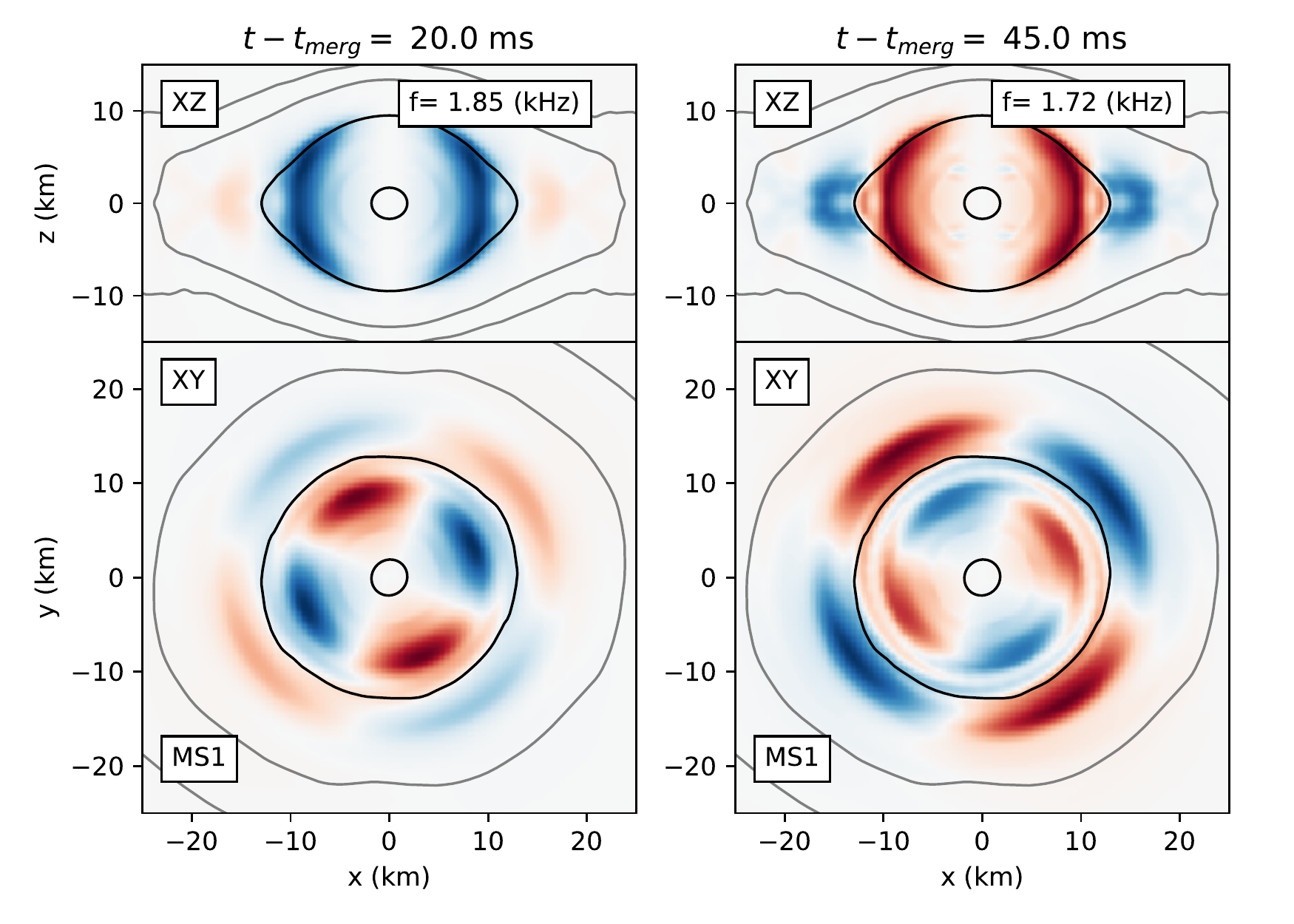}
\caption{As in Fig.~\ref{fig:APR4_mode_1}, but for EOS MS1 at \SI{20}{ms} and \SI{45}{ms} after merger.}
\label{fig:MS1_mode}
\end{figure}

\subsection{Mode eigenfunctions}
\label{subsec:wavefunction}

To analyze and show the difference of the various oscillation modes we
have performed the FFT transform in time segments of $5$ ms of the
density distribution and extracted the Fourier amplitude at selected
fixed frequencies (this correlates with the eigenfunction of a given
mode, as shown in \cite{2004MNRAS.352.1089S,Stergioulas:2011gd}). 
For each segment we  assign the central time as a label.  In Figures
\ref{fig:APR4_mode_1}, \ref{fig:APR4_mode_2}, \ref{fig:SLy_mode},
\ref{fig:H4_mode} and \ref{fig:MS1_mode} we summarize the behavior of
the density eigenfunction of each model for the frequencies we
mentioned in Section \ref{subsec:Spectrograms} at different
times. Each panel of those figures is normalized to the maximum value
of the density eigenfunction shown.  The first column of each figure
shows the eigenfunction of the dominant $m=2$ $f$-mode for each model
in both the equatorial (XY) and the vertical (XZ) planes. In the
equatorial plane there are no nodal lines, indicating that this is a
fundamental mode.
  
The APR4 EOS shows an excited mode already after $\sim$30 ms in the
postmerger phase, which is not visible in the spectrogram of
Figure~\ref{fig:MAIN} produced using the Prony's method, but is
clearly visible from the density eigenfunction profiles in Figure
\ref{fig:APR4_mode_1}. This is due to the fact that this excited mode
is extremely weak and cannot be distinguished using the Prony's
method. In Fig.~\ref{fig:MS1_mode} we note that the $m=2$ $f$-mode
for the MS1 does show a nodal line far from the center, in the
extended, low-density envelope. This behavior is not surprising for
differentially rotating stars in rapid rotation, where the
rotational forces can modify the structure of the eigenfuction in
the low-density regions near the surface, see
e.g. \cite{2004MNRAS.352.1089S}. 

The SLy EOS presents only an $m=2$ $f$-mode until $\sim$35 ms after
the merger, where we note the appearance of an excited mode at lower
frequency. The main mode is then suppressed and only the excited one
will survive until the remnant forms a black hole in the late post
merger phase. For the H4 EOS we note that the main $m=2$ $f$-mode is
the only mode present until $\sim\SI{45}{ms}$ when the excited mode at
lower frequency appears. The two modes then survive $\sim$30 ms until
the HMNS collapses to a black hole.

The excited low-frequency modes in the late postmerger phase have the
characteristics of inertial modes. Their frequency is in the right
range, with the dominant mode always having a frequency somewhat
smaller than the maximum rotational frequency in the star (although
our sample of models is small, it appears that the frequency of the
dominant mode in the late postmerger phase correlates with the
rotational frequency of the star, as expected from inertial
modes). Their eigenfunction has a peak in the low-density outer parts
of the stars, having more nodal lines than the fundamental quadrupole
mode. We note that inertial modes in rapidly rotating stars can be
axial-led (reducing to axial $r$-modes in the slow-rotation limit) or
polar-led (reducing to $g$-modes in the slow-rotation limit)
\cite{1999ApJ...521..764L,2003PhRvD..68l4010L}. The latter are also
called gravito-inertial modes. A precise identification of the
character of the excited modes would require a detailed comparison
with linear oscillations of quasi-equilibrium models. Because the
inertial modes observed in the postmerger phase are excited by
convective instabilities in the outer layers of the remnant, it is
likely that at least some (or most) of the inertial modes we observe
are polar-led, i.e. gravito-inertial modes (but the presence of
axial-led inertial modes cannot be excluded without further detailed
study).

\begin{figure*}
\includegraphics[width=\textwidth]{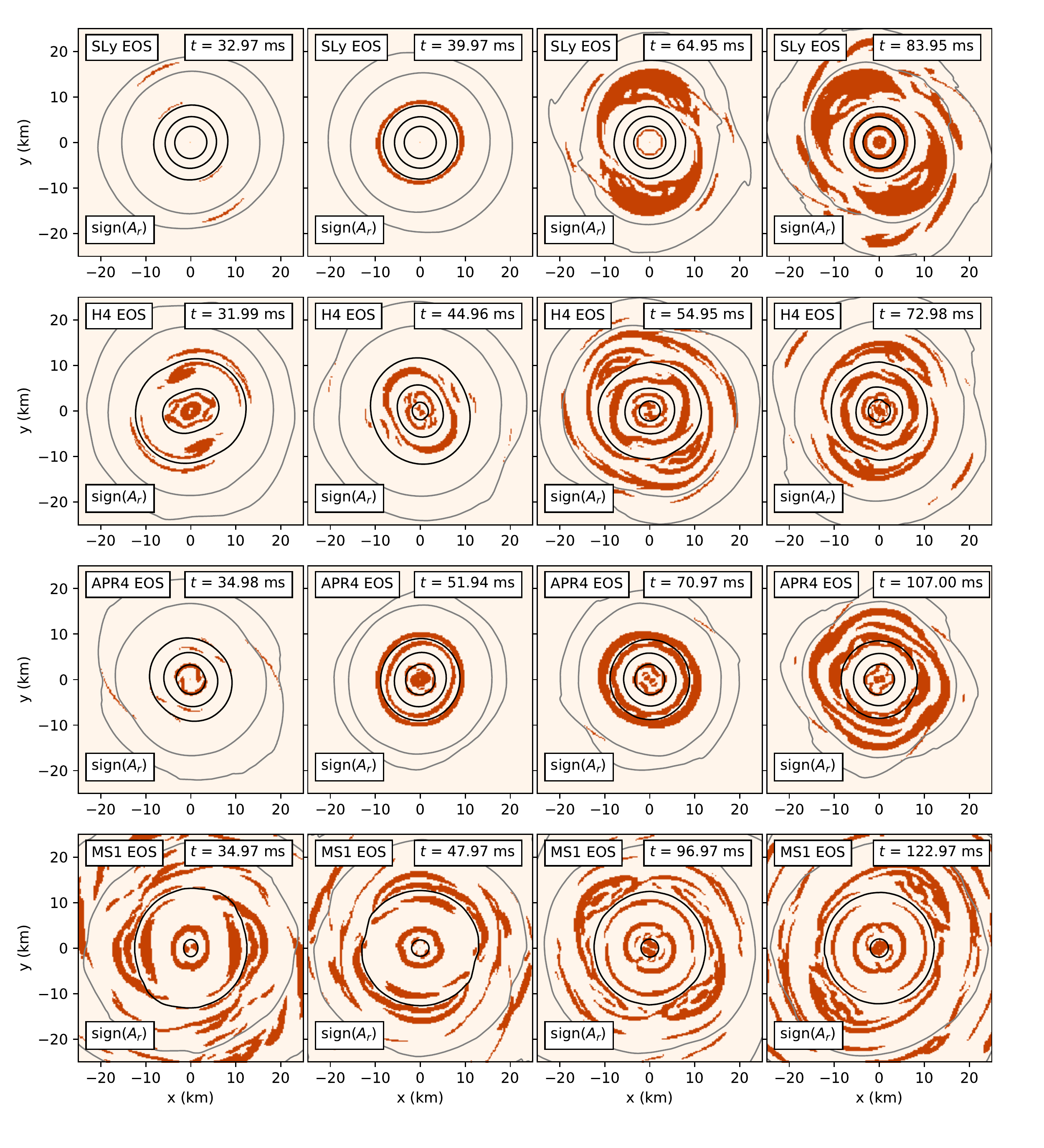}
\caption{Snapshots of $A_r$ in the equatorial plane. Each row
  corresponds to a different EOS model and each column to different
  times after merger. Dark color indicates regions where $A_r>0$,
  which corresponds to convective instability (see text for details). 
}\label{fig:Snapshots_Ar}
\end{figure*}

\begin{figure*}
\includegraphics[width=\textwidth]{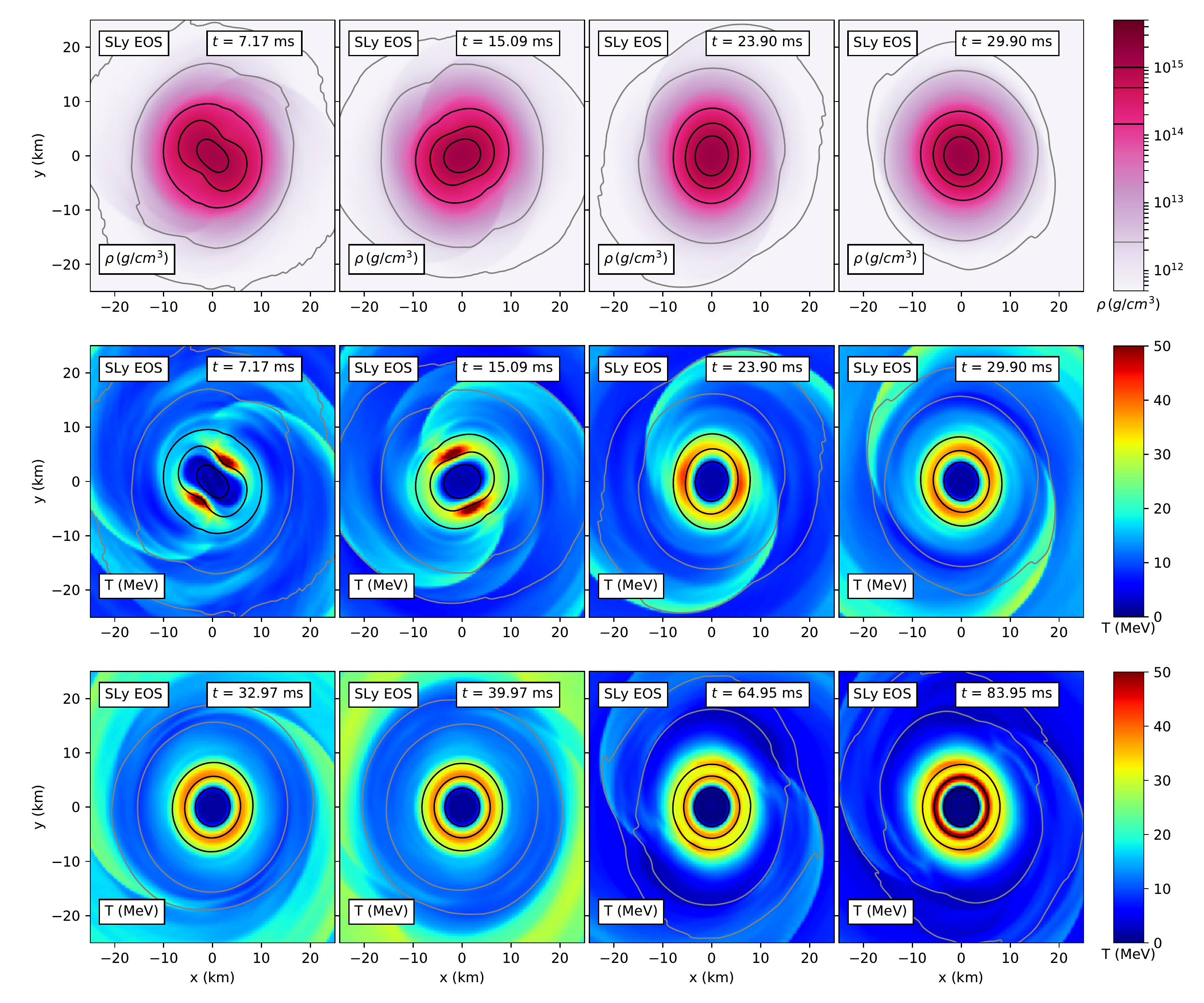}
\caption{Snapshots of the density and temperature in the equatorial
  plane for the the model with the SLy EOS at different times relative
  to the time of merger (see text for details). }
\label{fig:Snapshots_SLy}
\end{figure*}

\subsection{Convective instability and inertial modes}
\label{subsec:instability}

The appearance of lower-frequency modes in the late stage of our
simulations can be interpreted as inertial modes, for which the
Coriolis force is the dominant restoring force. The growth of such
modes is caused by a convective instability that appears in the
nonisentropic remnant just before the excited modes start to grow from
a small amplitude.

As we note in \cite{DePietri:2018}, the local convective stability depends on the 
Schwarzschild discriminant, which is defined as 
\begin{eqnarray}
A_\alpha = \frac{1}{\varepsilon+p}\nabla_\alpha \varepsilon - \frac{1}{\Gamma_1 p}\nabla_\alpha p \, ,
\end{eqnarray}
where
\begin{eqnarray}
\Gamma_1 
   := \frac{\varepsilon + p}{p} \left( \frac{dp}{d\varepsilon} \right)_s
    =         \left( \frac{d \ln p}{d \ln \rho} \right)_s
\end{eqnarray}
is the adiabatic index around a pseudo-barotropic equilibrium
and where $\varepsilon$ is the energy density. Regions where $A_\alpha < 0$ are 
convectively stable, while regions where $A_\alpha >0$ are convectively {\it unstable}.
Since in this work we are using a piecewise polytropic approximation for the
construction of our models, we calculate $\Gamma_1$ as
\begin{eqnarray}
\Gamma_1 = \Gamma_{\text{th}} + (\Gamma_i - \Gamma_{\text{th}}) \frac{K_i \rho^{\Gamma_i}}{p}
\end{eqnarray}
where $K_i$ and $\Gamma_i$ are the polytropic constant and exponent in
the $i$-th piece of the EOS, respectively, and $\Gamma_{\text{th}}$ is
the adiabatic index added for the thermal treatment (see Appendix
\ref{app:EOS} for further details). In order to determine the
convective stability in the long-living remnant, we calculate $A_r$
and $A_\theta$ in the equatorial and vertical planes at different
times.

The results of the computation of the Schwarzschild discriminant in the
equatorial plane for each model considered in this work
are displayed in Fig.~\ref{fig:Snapshots_Ar} at selected evolution times.
In this figure convectively unstable regions ($A_r > 0$) are represented using
a dark color.  The first row of Fig.~\ref{fig:Snapshots_Ar} shows the
case of the SLy EOS and we note that the remnant is stable in the
first $\sim$35 ms of the postmerger phase and the GW spectrum is
dominated by the $m=2$ $f$-mode. After $\sim$39 ms we observe the
formation of a convectively unstable ring, which corresponds to the
outer layer of the hot ring which is visible in the second panel of
the third row of Fig.~\ref{fig:Snapshots_SLy}, depicting temperature isocontours. 
This coincides with the
first appearance of inertial modes. At $\sim$65 ms the convectively
unstable ring has expanded to the lower density regions of the remnant and appears
fragmented, which correspond to the strong growth of an inertial mode
at $2.38$ kHz. The amplitude of the mode remains almost constant
throughout the following phase, while the convectively unstable
regions keep expanding in the envelope until the remnant finally collapses to a
black hole.

In the second row of Fig.~\ref{fig:Snapshots_Ar} we display the Schwarzschild discriminant
for the H4 EOS, which presents a slightly different situation compared to the SLy
EOS. Around $\sim$32 ms we note that there are unstable regions
in the outer layer of the core which are due to the presence of
hot spots (not shown in this work) generated by the
shock heating produced during the merger. At $\sim$45 ms an unstable ring appears
just outside the core, corresponding to the growth of an inertial mode
with frequency $2.11$ kHz. The unstable regions expand and become fragmented at a
later time whereas the amplitude of the inertial mode continues to grow until the remnant 
collapses to a black hole after $\sim$84 ms.

The evolution of the Schwarzschild discriminant for the model with the APR4 EOS, which is shown in the third row of
Fig.~\ref{fig:Snapshots_Ar}, appears at first glance to be similar to that of the
SLy EOS. However, it does show a crucial difference, visible in the first panel of the figure. This panel
displays a generally stable remnant despite, as we have already
observed in Sec.~\ref{subsec:wavefunction}, the presence of an additional
excited mode by this evolution time ($\sim$35 ms)). The appearance of this mode is not
triggered by the expansion of the convectively unstable region, since
in this phase the remnant has a corotating region which is visible in
Figure~\ref{fig:APR4rot}.  This figure shows that the mode
frequency is lower than the peak of the rotational profile at $t=25$ ms,
which corresponds to the maximum rotation frequency for the APR4 EOS model.
As in the case of the SLy EOS, a convectively unstable ring forms at a
later time, a few milliseconds before the appearance of inertial
modes. The unstable regions starts expanding in the lower density
regions of the remnant, leading to the growth in amplitude of such
modes.

Finally, in the case of the MS1 EOS, the different snapshots in the last row
of Fig.~\ref{fig:Snapshots_Ar} show convectively unstable regions fairly spread over the
entire postmerger remnant from earlier times than for the other three models.
Notice that for this EOS, the episodes of
excitation-saturation-destruction of a lower-frequency inertial mode
already start at earlier times than shown in Figure
\ref{fig:Snapshots_Ar}.

Fig.~\ref{fig:Snapshots_SLy} shows snapshots of the density and
temperature in the equatorial plane for the model with the SLy EOS, at
different times relative to the time of merger. In the first two rows
of the figure we display the density and temperature profiles of the
postmerger remnant, up to a time of \SI{29}{ms}, when the $m=2$
$f$-mode is still dominant. We note that in this phase there are two
hot spots where the remnant is deformed and which start to
diffuse. This is the expected behavior, as found in simulations with
many different numerical codes (see e.g.~\cite{2019arXiv190708534B}
and references therein). In the last row of the figure we show
snapshots of only the temperature at times between \SI{33}{ms} and
\SI{84}{ms}. A hot shell forms around the cold core at $\sim$33
ms. Later, at around \SI{65}{ms}, a second hot ring is forming in the
equatorial plane and by \SI{84}{ms} the temperature of the inner shell
has risen visibly, reaching about $T\sim 50$ MeV.

In our simulations, convection is exciting specific, global inertial
modes. This is not unexpected, see
e.g.~\cite{Zhang:1987,Simitev:2003,Busse:2004,Zhangbook:2017}. We note
that the inertial modes are excited when the temperature profile in
the remnant has become nearly axisymmetric (by that time, the
amplitude of the peak $m=2$ f-mode oscillation has diminished). In
contrast, the initial strongly nonaxisymmetric temperature pattern
(essentially in the form of two hot spots) does not seem to lead to
the excitation of global modes. This is true for all four models
considered: inertial modes start growing once the temperature
distribution has become sufficiently circular in the equatorial plane.

We stress that the convective instability found in our simulations is
solely triggered by the thermal gradient created by shock heating at
the time of merger. Hence, in our setup no additional physics
(e.g.~radiative cooling) is necessary to create a thermal gradient.
As shown in Fig.~\ref{fig:Snapshots_SLy}, for $t>30$ ms after merger,
the core of the remnant is cold and is surrounded by a shell of hot
matter, with the temperature reaching a peak at about 5 km from the
center. Beyond that distance, the temperature decreases steeply with
radius and it is this thermal gradient that is responsible for
triggering the convective instability. A similar situation has been
recently reported by~\cite{Camelio:2019} who study rotating stars with
non-barotropic thermal profiles. This work shows that convection sets
in without requiring additional physics, as long as there is a
negative entropy gradient when the density is below a critical density
(see Eq.~C7 in~\cite{Camelio:2019}), which is the case in our
simulations.

In particular, numerical dissipation does not play a role in
triggering the convective instability. It can only dampen the
instability, once it sets in, or entirely prevent it, if it is too
strong. In our simulations, the numerical dissipation is sufficiently
small to allow for the instability to grow and saturate.
Fig.~\ref{fig:HamConstr} in Appendix~\ref{app:NM} displays the
evolution of different norms of the Hamiltonian constraint violations,
at three different resolutions (for EOS Sly as a representative
case). The violations converge with resolution and remain sufficiently
small throughout the duration of the simulation.

\begin{figure}
\includegraphics[width=0.45\textwidth]{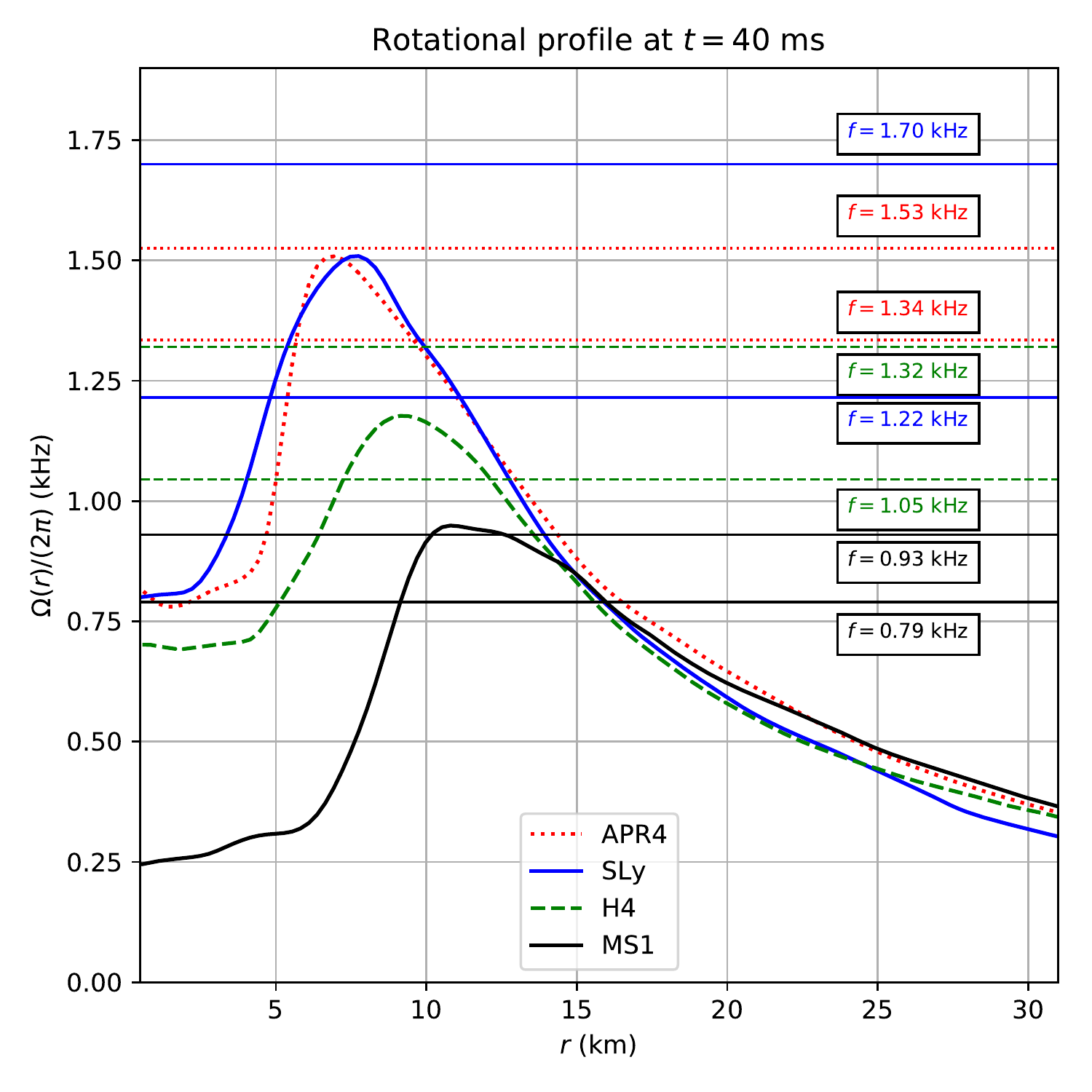}
\vspace{-4mm}
\caption{{Rotational profiles $\Omega/2\pi$ vs. coordinate radius $r$,
    for the four different EOS models at $t=40$ ms after merger. The
    horizontal lines indicate the pattern speed frequencies of the two main $m=2$ modes for each model. 
    The colors of the horizontal lines are the same as for the
    EOS labels. } }
\label{fig:Rotation40}
\end{figure}

\begin{figure}
\includegraphics[width=0.45\textwidth]{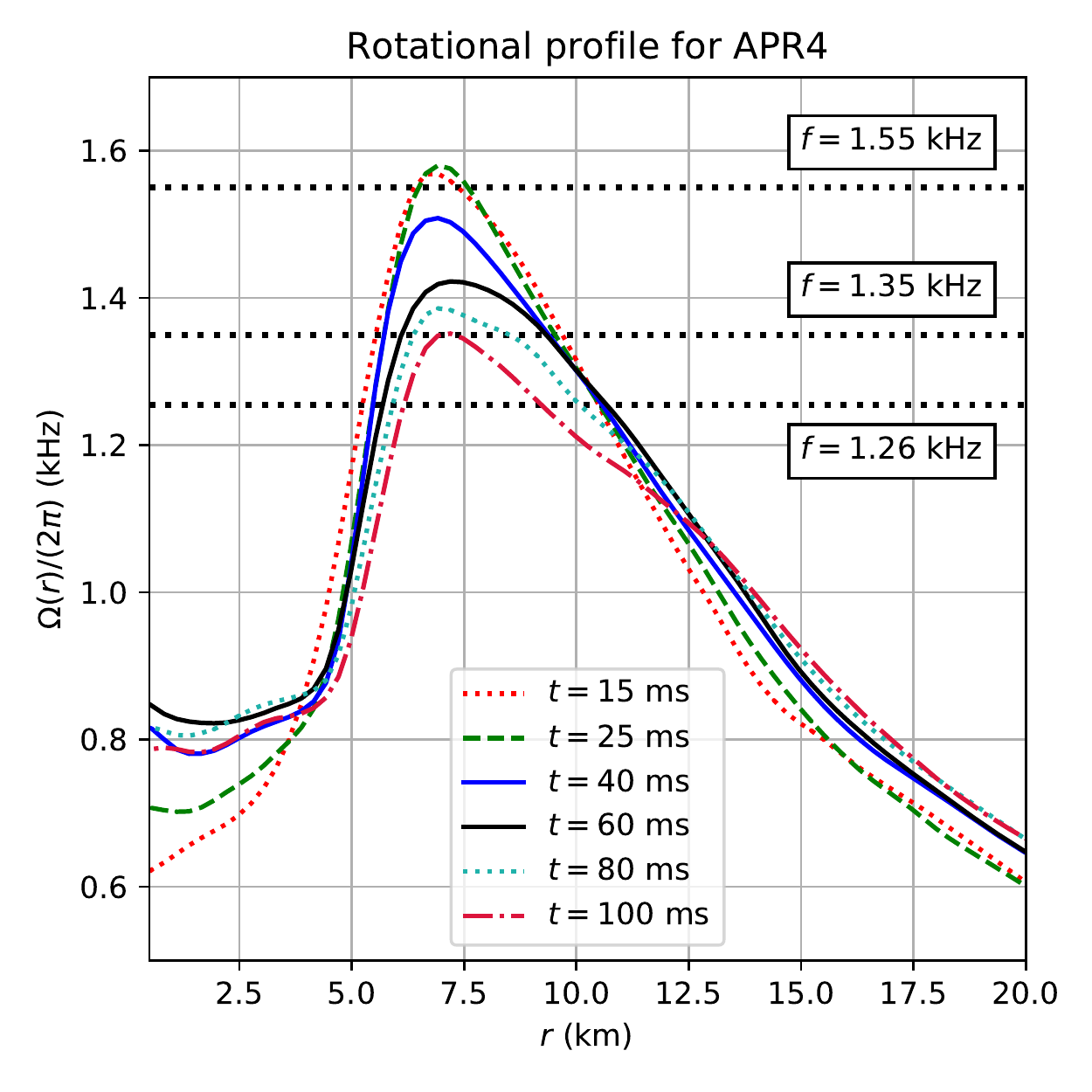}
\vspace{-4mm}
\caption{{Rotational profiles $\Omega/2\pi$ vs. coordinate radius $r$,
    for the model with EOS APR4 models at different times after
    merger.  The horizontal lines indicate the pattern speed (assuming
    $m=2$) of the $f$-mode (\SI{1.55}{kHz}) and of the main inertial mode at
    the start (\SI{1.35}{kHz}) and the at end (\SI{1.26}{kHz}) of the convectively
    unstable phase. }}
\label{fig:APR4rot}
\end{figure}

\subsection{Corotation and mode revival}
\label{subsec:corot}

As noted in the introduction, the state of the system in the
postmerger phase is a single (excited) differentially rotating
neutron star with a non trivial rotational profile, 
{\bf which is shown in Figures~\ref{fig:Rotation40} and \ref{fig:APR4rot}.
Moreover, there is also an non-trivial temperature profile
(and indeed a non-trivial entropy)  profile, in which the temperature has an off-center
maximum (a circle in the equatorial plane at late times).
The rotational and thermal structure of the postmerger remnants in our simulations is in agrement with results in the literature  (see~e.g. \cite{Shibata:2006nm,Hanauske:2016gia,Kastaun:2016yaf}) where similar assumptions were made
(only when assuming a very  strong physical viscosity different profiles are found, see~\cite{2017PhRvD..95l3003S}).
}
The remnant may be subject to various instabilities, namely the dynamical bar-mode
instability (see
\cite{Baiotti:2006wn,DePietri:2014mea,Loffler:2014jma} and references
therein), the secular Chandrasekhar–Friedman–Schutz (CFS)
instability \cite{PhysRevLett.24.611,1978ApJ...222..281F}
and shear instabilities that can develop in differentially rotating 
stars (see e.g. \cite{Corvino:2010} and references therein).
In addition, our results reveal the possibility
of the excitation of additional modes in the postmerger phase for
sufficiently long times. We discuss next how the frequency of the
excited inertial mode is related to the rotational profile of the
neutron star at the late stage of the postmerger dynamics.

In Fig.~\ref{fig:Rotation40} we show the rotational profile (angular
velocity vs. distance in the equatorial plane) of the four models at
$\SI{40}{ms}$ after merger, along with the pattern speed frequency for the two main
modes in each model, indicated by the horizontal lines. For $m$=2 modes the pattern 
speed frequency is half of the associated GW
frequency. We note that at this stage of the evolution only the 
APR4 and MS1 models are still in corotation with the mode pattern
speed.

In Figure~\ref{fig:APR4rot}, we focus on the model with the APR4 EOS
and show the rotational profile at different times, along with the
pattern speed of the $m=2$ $f$-mode (\SI{1.55}{kHz}) and of the main
inertial mode at the start of its excitation (\SI{1.35}{kHz}) and at the
end of the simulation (\SI{1.26}{kHz}). It is interesting to notice
that the revival of the amplitude of the $m=2$ $f$-mode at \SI{25}{ms} 
after merger coincides with the mode being in corotation with the 
HMNS in a small region. This process is possible when the corotation
of the star occurs at the same time as the convective instability 
of a small region of the star. In this case, as very small regions of the remnant
present $A_r > 0$ at \SI{25}{ms}, angular momentum from the
rotation of the star is injected into the main $f$-mode.

\section{Conclusions}
\label{sec:conclusions}

We have presented the results of new numerical simulations of BNS systems
in full general relativity, where the simulations have been extended
up to $\sim$140 ms after merger.  In addition to the results for models based on the SLy and APR4 EOS
we already discussed in~\cite{DePietri:2018}, and whose analysis has been significantly extended here, 
we have also studied models equipped with two more EOS, namely H4 and MS1 EOS. The main focus of this work has been the
analysis of the spectrum of the postmerger GW signal. Our investigation has been driven, in particular, by our aim to provide
further evidence in support of our claim in~\cite{DePietri:2018} about the potential existence of new families
of oscillation modes excited in long-lived postmerger remnants (in addition to the main fluid quadrupole $f$-mode) at sufficiently late times of
the evolution.

In~\cite{DePietri:2018} we already observed that the appearance of
convectively unstable regions and the excitation of inertial modes in
remnants that survive for a long time after merger depend on
the rotational and thermal state of the remnant and affect the HMNS
dynamical evolution. In all further cases investigated in the present work, the 
late-time excitation of inertial modes to large amplitudes has been corroborated. 
Indeed, inertial modes become the dominant modes for the GW emission in the
postmerger phase of pulsating HMNS for times from about \SI{40}{ms} up to 
at least \SI{100}{ms} after merger. We have found that the postmerger phase 
can be subdivided into three phases: an early postmerger phase (where the 
quadrupole mode and a few subdominant features are active), the intermediate 
postmerger phase (where only the quadrupole mode is active) and the late 
postmerger phase (where convective instabilities trigger inertial modes).
For all of the models of our sample, the impact of the inertial modes in the 
GW spectrum is significant, appearing at signal-to-noise ratios of immediate 
interest for 3rd-generation detectors, such as the Cosmic Explorer~\cite{Evans:2016mbw} 
and the Einstein Telescope~\cite{Punturo:2010zz}. This allows for the possibility 
of probing not only the cold part of the EOS but also its dependence on finite temperature. 

Recently~\cite{Camelio:2019} have analyzed convectively unstable
rotating neutron stars with non-barotropic thermal profiles (as in the
case of binary neutron star remnants) reporting a growth timescale of
convection of ${\cal O}(10)$ ms close to the center and ${\cal
  O}(0.1)$ ms at lower densities (close to the surface). These
timescales are compatible with those presented in~\cite{DePietri:2018}
and further corroborated in this work. Our study has neglected the
effects of shear and bulk viscosity in the star as well as the
effective viscosity due to MHD turbulence. In view of our results, it
will be important to investigate the influence of viscosity on the
lifetime of the remnant and their impact on the rotational profiles of
the HMNS, as this might affect the potential excitation of inertial
modes. Recent work by \cite{2017PhRvD..95l3003S,2018ApJ...860...64F},
in which effective shear viscosity was modelled through the so-called
viscous $\alpha$-parameter, has shown that the degree of differential
rotation in the postmerger remnant is significantly reduced in the
viscous timescale, $\le 5$ ms, for $\alpha\sim{\cal O}(10^{-2})$. It
remains to be found if such (high) values of the $\alpha$-parameter
are reached in MHD simulations of BNS mergers involving pulsar-like
magnetic-field strengths in the inspiral phase.
    
\acknowledgments

This project greatly benefited from the availability of public
software that enabled us to conduct all simulations, namely ``LORENE''
and the ``Einstein Toolkit''. We express our gratitude to everyone
that contributed to their development.  We have benefited from
discussions with Alessandro Drago, Giuseppe Pagliara, Silvia Traversi
and Masaru Shibata and we are very grateful to them. We acknowledge
PRACE for awarding us access to MARCONI at CINECA, Italy, under grant
Pra14\_3593.  This work also used resources provided by the
CINECA-INFN agreement that provides access to GALILEO and MARCONI at
CINECA and by the Louisiana Optical Network Initiative (QB2,
allocations loni\_hyrel, loni\_numrel, and loni\_cactus), and by the
LSU HPC facilities (SuperMuc, allocation hpc\_hyrel).
JAF acknowledges financial support provided by the
Spanish Agencia Estatal de Investigaci\'on (grants
AYA2015-66899-C2-1-P and PGC2018-095984-B-I00), by the Generalitat
Valenciana (PROMETEO/2019/071) and by the European Union’s Horizon
2020 research and innovation (RISE) programme H2020-MSCA-RISE-2017
Grant No. FunFiCO-777740.  FL is directly supported by, and this
project heavily used infrastructure developed using support from, the
National Science Foundation in the USA (Grants No. 1550551,
No. 1550461, No. 1550436, No. 1550514).  NS is supported by the ARIS 
facility of GRNET in Athens (GWAVES and GRAVASYM allocations)'' Support from INFN
``Iniziativa Specifica NEUMATT'' and by the COST Actions MP1304
(``NewCompStar'') and CA16104 (``GWVerse''), is also kindly
acknowledged.

\appendix

\section{Comparing simulations at different resolutions }
\label{app:NM}

\begin{figure}
\includegraphics[width=0.45\textwidth]{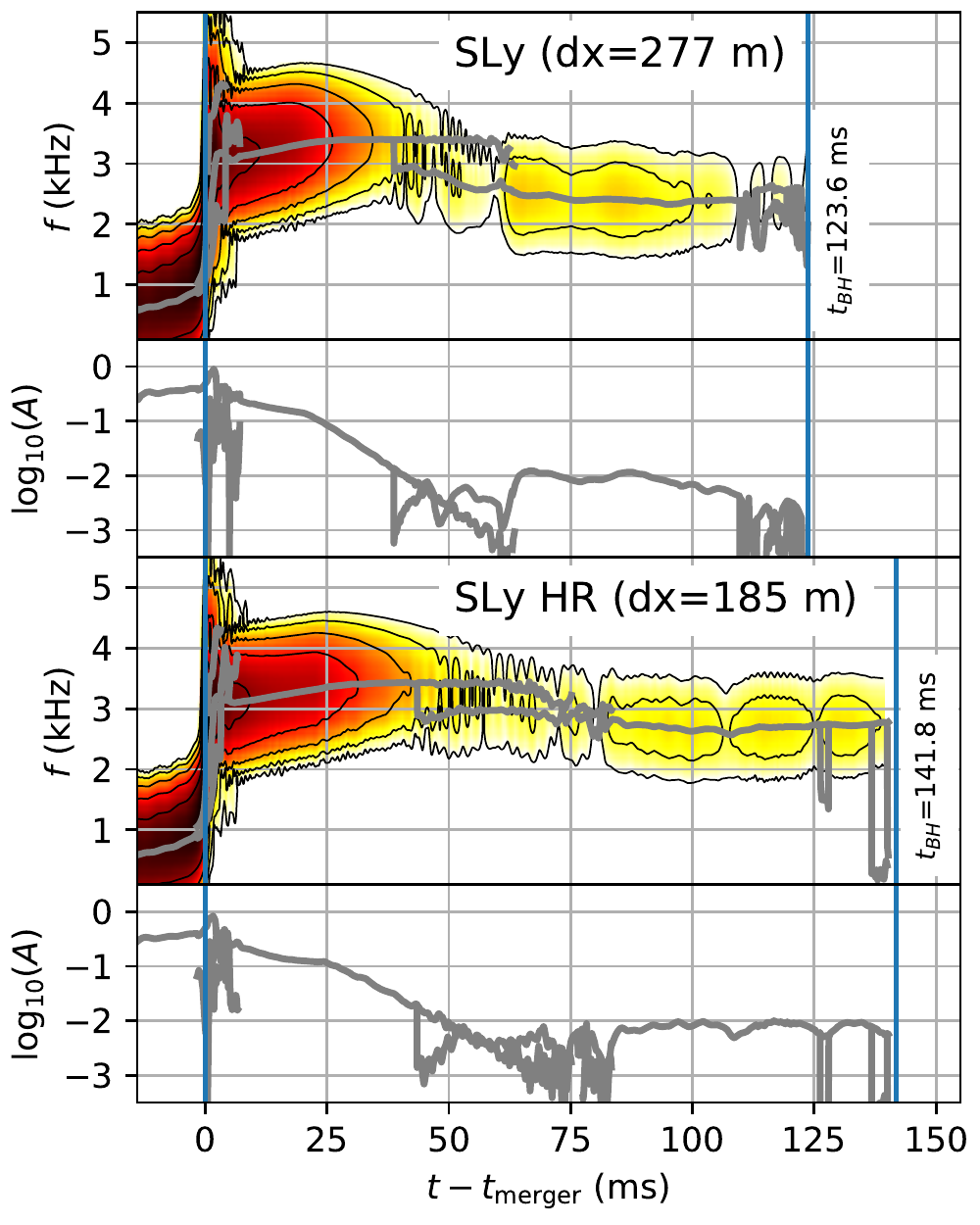}
\vspace{-4mm}
\caption{{Time-frequency analysis for the model with EOS SLy at two different resolutions (see text for details).}}
\label{fig:SLy_12res}
\end{figure}

The convergence properties of the numerical code and the accuracy of
the simulations were discussed in \cite{DePietri:2015lya}.  Here, we
compare the long-term evolution of the model with EOS\ SLy using three
different resolutions of the innermost grid, namely $dx=\SI{369}{m}$,
$dx=\SI{277}{m}$ (the standard resolution in this work), and
$dx=\SI{185}{m}$ (which is about five times more computationally
expensive).

\begin{figure}
\includegraphics[width=0.45\textwidth]{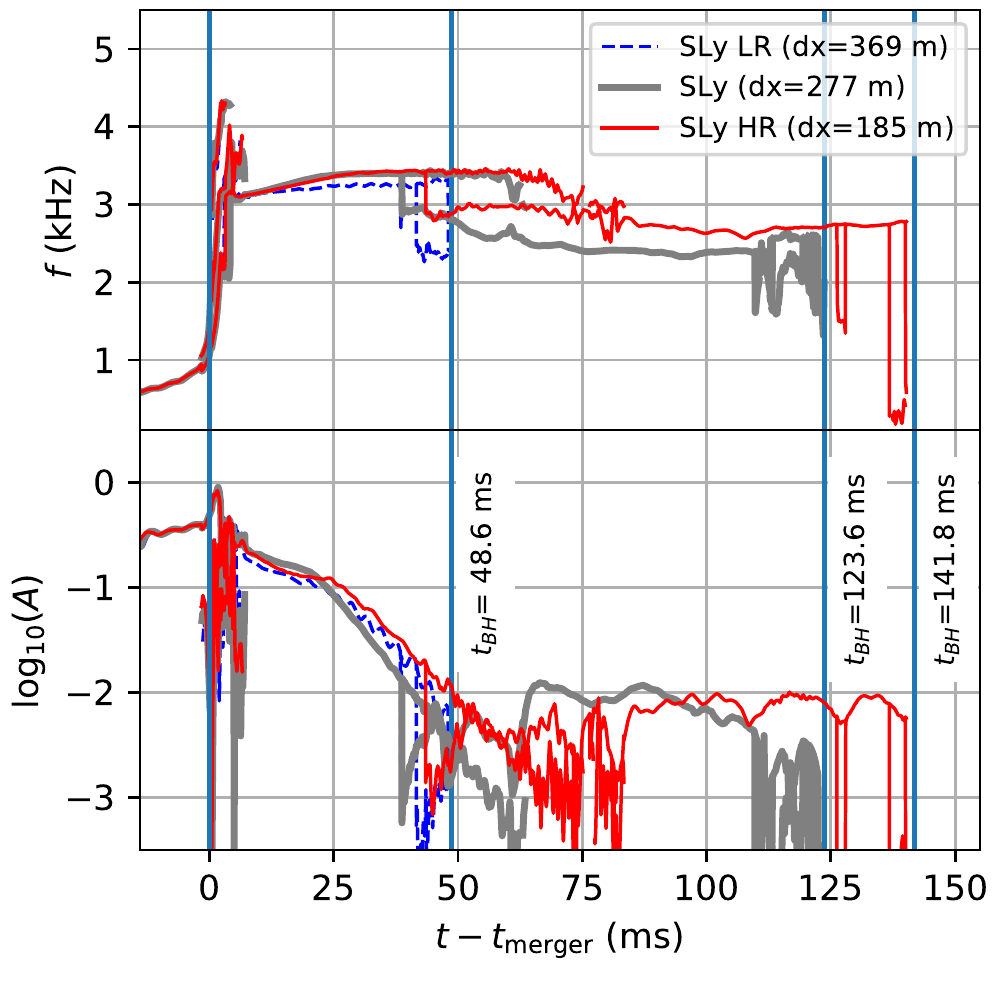}
\vspace{-4mm}
\caption{Comparison of the Prony analysis for the model with EOS SLy at three different
resolutions.}
\label{fig:App3res}
\end{figure}

Figure 14 shows the spectrogram and the Prony's analysis for the
two highest resolutions, while Figure~\ref{fig:App3res} shows the comparison of
results of the Prony's analysis for all the three considered
resolutions. From the data shown in these figures we conclude that
the appearance of the convective instabilities, and the associated
excitation of inertial modes, happens at somewhat later times as the
resolution is increased. Likewise, black hole formation is delayed in
all three cases, taking longer times for higher resolutions, as
discussed in~\cite{DePietri:2015lya}. Also, there are three (instead
of two) episodes of excitation-saturation-destruction before the final
excitation of a long-lasting inertial mode with stable saturation
amplitude. At the lowest resolution, the appearance of an inertial
mode (dashed blue line in Fig.~\ref{fig:App3res}) is rapidly
suppressed due to the collapse of the HMNS to a black hole. For the
standard and high-resolution simulations this mode survives for a
longer time due to the smaller numerical dissipation at higher
resolutions. For a resolution of $dx=\SI{185}{m}$ the inertial mode
appears somewhat later and at a higher frequency than for
$dx=\SI{277}{m}$ and $dx=\SI{369}{m}$ since the dissipation at such
resolution is smaller and leads to a slightly different hydrodynamical
evolution.  However, the frequency of the main $f$-mode does not
depend strongly on the resolution.

Whereas we do observe an overall delay in the appearance of various
dynamical features in the simulations (which could be due to a better
preservation of angular momentum at higher resolution) the qualitative
features between the three resolutions remain the same.  The results
of Figure \ref{fig:App3res} seem to contradict the assertion
in~\cite{Breschi:2019srl} that some features in the frequency spectrum
reported in~\cite{DePietri:2018} and also here disappear using higher
resolution.  We expect that this discrepancy can be resolved by a
closer examination of the numerical data in \cite{Breschi:2019srl}, if
these become available (or by additional high-resolution
simulations). We further comment on this issue in Appendix
\ref{sec:relatedworks}.

\begin{figure}
\includegraphics[width=0.45\textwidth]{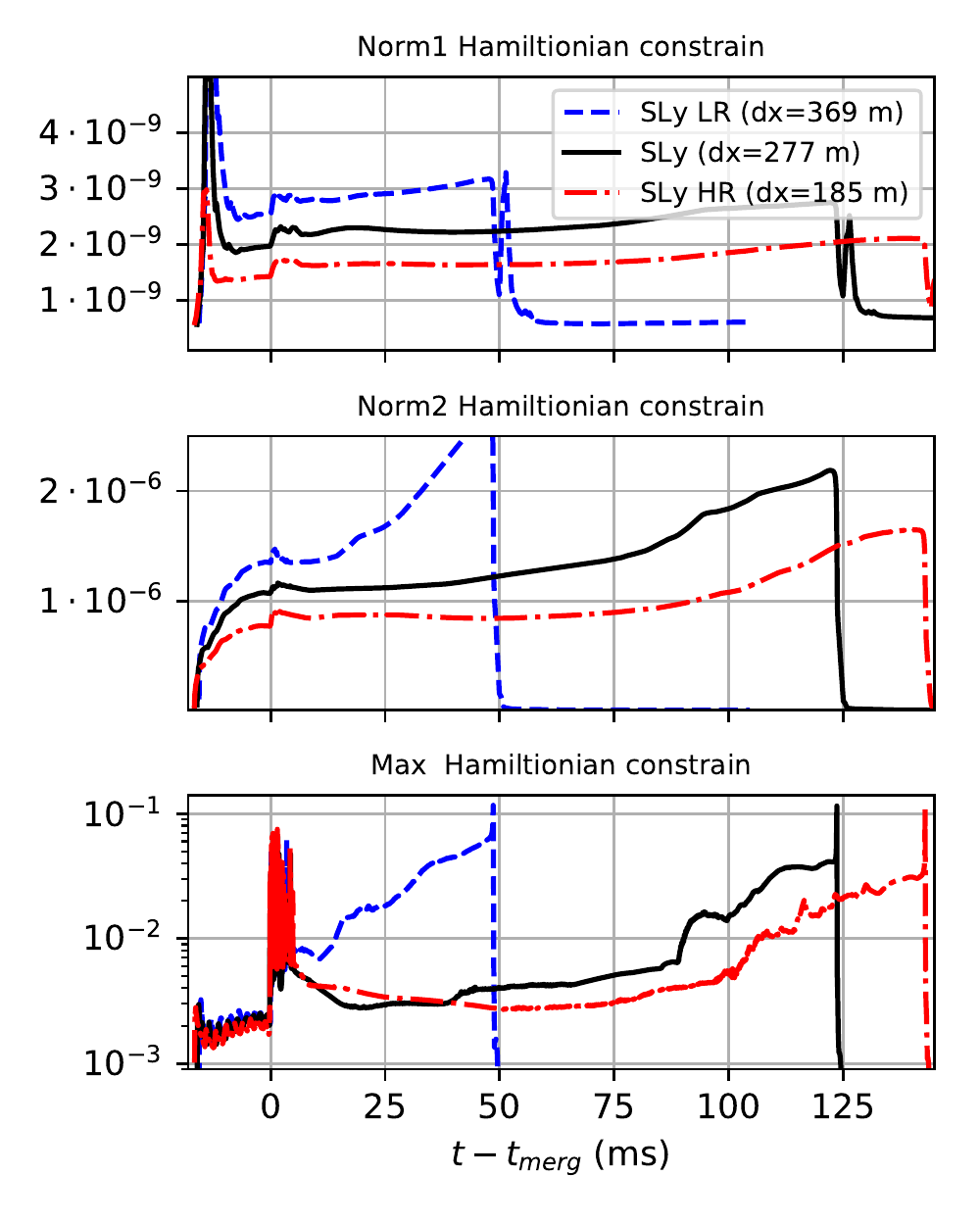}
\vspace{-4mm}
\caption{{Hamiltonian constraint violation for different resolutions of the model with EOS SLy.}}
\label{fig:HamConstr}
\end{figure}

Figure \ref{fig:HamConstr} depicts the violation of the Hamiltonian
constraint $\mathcal{H}(x)$ for the simulation of the model with EOS
SLy as a representative case and at the same three numerical
resolutions.  The top panel shows the $L_{1}$ norm (i.e.~the
coordinate integration over the whole grid of $|\mathcal{H}(x)|$), the
middle panel depicts the $L_{2}$ norm (the corresponding integration
of $|\mathcal{H}(x)|^2$) and the bottom panel shows the maximum of the
constraint violation of $\mathcal{H}(x)$ over the whole grid. The
average quantities converge when increasing the grid resolution. The
same holds true for the maximum value of $\mathcal{H}(x)$ except for
the short time in which the two neutron stars merge. The largest
constraint violations are attained at black hole formation, as usual
in this type of simulations. Overall, this figure shows that the
simulations at different resolutions are consistent and that the
violations converge with resolution and remain sufficiently small
throughout the duration of the simulation.

\section{Notes on previous results for simulations lasting 100 ms after merger.}
\label{sec:relatedworks}

Our study is not the first one in which the postmerger phase of a
binary neutron star merger is evolved for timescales of about
\SI{100}{ms}. Previous works, all of them fairly recent,
include~\cite{Rezzolla:2010fd,Andersson:2013mrx,Zhang:2017,Ciolfi:2019fie,Breschi:2019srl}.
Because of the very small amplitude of the GWs emitted in the late
convective phase (as opposed to the large amplitude emitted by
hydrodynamical motions during the early phase) one can easily miss the
presence of convectively-excited modes in the late postmerger phase,
or one can simply dismiss these as artifacts, without further
analysis. To the best of our knowledge, other groups have not
specifically examined long-term simulations in terms of eigenfunctions
or indicators of convective instability.  The studies
of~\cite{Rezzolla:2010fd} and~\cite{Andersson:2013mrx} report the
results of simulations of binary neutron star mergers with initial
data constructed with a simple polytropic $\Gamma=2$ EOS and evolved
as an ideal fluid with $\Gamma_\mathrm{th}=2$, extending up to
$\SI{~140}{ms}$.  Similarly long-lasting simulations for the same EOS
are discussed in~\cite{Zhang:2017}. None of those investigations
report the late-time excitation of inertial modes in the remnant. This
indicates that such modes are likely not easily excited when the
initial data and the ideal fluid EOS have the same index $\Gamma$
(notice e.g. that in \cite{Rezzolla:2010fd} that main $l=m=2$ $f$-mode
still has a large amplitude at the end of the simulation, whereas in
our simulations the convective instabilities always set in only after
the amplitude of $l=m=2$ $f$-mode has diminished).

The work of~\cite{Ciolfi:2019fie} shows the presence of a second mode
at late postmerger times.  In the top panel of their Figure 16 they
show the amplitude of the $h_{22}$ GW-mode for a simulation of a
long-lived magnetized neutron star merger described by the APR4 EOS
and masses $1.35+1.35$$M_\odot$. This model is slightly more massive
than the one with the same EOS used in this work. The simulation is
followed up to $\approx \SI{100}{ms}$ after merger. This figure shows
the presence of an unphysical gauge mode with a frequency of $\approx$
\SI{1}{kHz} around \SI{70}{ms} after merger. On the other hand the
spectrogram shown in the bottom panel of their figure 16 displays the
presence of a single mode up to \SI{30}{ms} after merger while two
modes become visible at a later time. This suggests a similar
situation to the one described in the present work. Indeed,
while~\cite{Ciolfi:2019fie} do not arrive at an explanation for this
second frequency, they notice the similarity to our results
in~\cite{DePietri:2018}. We note, however, that an unambiguous
comparison would only be possible by extracting the associated
eigenfunctions of the late-time modes found by~\cite{Ciolfi:2019fie}.

Appendix B of~\cite{Breschi:2019srl} presents a study of convergence
for a model very similar to the one discussed in the present work,
namely a model with EOS SLy4 and masses $1.30+1.30$$M_\odot$. We
recall that here we have considered a model described by the
piecewise-polytropic representation of the SLy EOS with masses
$1.28+1.28$$M_\odot$. The resolutions used by~\cite{Breschi:2019srl}
are $dx=\SI{415}{m}$ (VLR), $\SI{246}{m}$ (LR), $\SI{185}{m}$ (SR),
and $\SI{136}{m}$ (HR).  The GW strain and time-frequency diagram are
shown in their Figure 13. In the postmerger phase the convergence
properties of their simulations are poor (they had to use a rather
small Courant-Friedrichs-Lewy factor of 0.075) with the VLR and SR
simulations behaving differently to the LR run. Moreover, their HR
simulation shows the formation of a black hole within less that
\SI{30}{ms} after merger (while this does not happen at the other
resolutions) and that the phase of the waveform is not in the
convergent regime. Nevertheless, the visual inspection of their
results at VLR and SR resolutions seem compatible with our
results. The discrepancy might be resolved by a closer examination of
the numerical data of~\cite{Breschi:2019srl}.

Finally~\cite{Camelio:2019} present a study of convectively unstable
rotating neutron stars with non-barotropic thermal profiles (as in the
case of binary neutron star remnants). In their simulations, the
thermal gradients naturally trigger convective motions in the star and
the authors estimate (see their equation C8) that the growth timescale
of convection is of order tens of ms close to the center, but reduces
to the order of 0.1 ms at lower densities (close to the surface in
their models). As~\cite{Camelio:2019} note referring to the rapid
growth of convective instabilities in our simulations in the
low-density envelope of the remnant, these timescales are compatible
with those we find in~\cite{DePietri:2018} (and here). This is
additional evidence in favor of our explanation of the late
postmerger phase in our simulations.

\section{Piecewise polytropic EOS}
\label{app:EOS}

The numerical simulation of gravitational systems that include neutron
stars requires a description of the behavior of matter at the very
high density involved inside such compact objects.  A full description
would require a detailed understanding of the properties of matter at
densities exceeding nuclear density, which is not yet fully
accomplished.  For this reason and for computational convenience, we
here choose to model thermal effects similar to an ideal fluid. In
this approximation, the energy momentum tensor is written as
$T^{\mu\nu} = \left[\rho (1+\epsilon) + p\right] u^\mu u^\nu+p
g^{\mu\nu}$ where $\rho$ is the baryon density, $\epsilon$ is the
specific internal energy and $p$ is the pressure. As a consequence,
the energy density is $e=\rho (1+\epsilon)$. At this point, the
thermal description of the system is achieved assuming (where
$\epsilon_{\rm th}$ is an arbitrary function of thermodynamical state
that has the properties of being zero at zero temperature):
\begin{eqnarray}
\epsilon &=& \epsilon_0(\rho) + \epsilon_{\rm th}, \\
p        &=& p_0(\rho) + (\Gamma_{\rm th} -1 ) \rho \epsilon_{\rm th}, 
\end{eqnarray}  
and $\epsilon_0(\rho)$ and $p_0(\rho)$ are the specific internal energy and
 pressure at $T=0$. For a fixed number of fermions the volume is  $V= (1/\rho)$, 
and the energy $E=e/\rho$. The first law of thermodynamics is:
\begin{equation}
 dQ =  dE + p dV = T dS.
\label{eq:dq}
\end{equation}
Assuming thermodynamic consistency, Eq.~(\ref{eq:dq}), Frobenius theorem uniquely fixes (up to 
an arbitrary function, i.e., any function $\Sigma(s)$ gives other 
possible definitions of ``temperature'' and ``entropy'' (s,t) as $S=\Sigma(s)$ and 
$T=t/(d\Sigma(s)/ds)$) the ``temperature''-like and ``entropy''-like function as:
\begin{eqnarray}
   T &=& (\Gamma_\mathrm{th}-1) \epsilon_{\rm th}, 
\label{eq:temp}\\
   S &=& \log \left[
      \frac{  (\epsilon_{\rm th})^{1/(\Gamma_\mathrm{th}-1)}}{
              \alpha\rho}  \right],
\label{eq:entropy}
\end{eqnarray}
where $\alpha$ is an arbitrary constant with dimensions of
the inverse of density. The conversion factor from the above dimensionless
``temperature''-like values $\epsilon_{\rm th}$ and the real
temperature $T$ is given by the specific heat at constant volume and
the function $\Sigma(s)$.  The choice of ($\Sigma(s)=s$) corresponds
to having an ``ideal-fluid'' thermal behavior, and, as discussed
in~\cite{rezzolla2013relativistic}, the dimension of the temperature
$T$ can be found by multiplying Eq.~(\ref{eq:temp}) by a factor
$m/k_B$, where $m$ is the value of the baryon mass.  At the same time,
the condition of thermodynamic consistency at $T=0$ determines the
functional relation between $\epsilon_0(\rho)$ and $p_0(\rho)$ ($dQ=0$
at $T=0$) that reads:
\begin{equation}
 p_0(\rho) = \rho^2 \frac{d\epsilon_0(\rho)}{d\rho}.
\label{eq:PP}
\end{equation}
The zero-temperature piecewise polytropic approximate EOS amounts to the 
assumption that  $p_0(\rho)$ and $\epsilon_0(\rho)$ are continuous 
polytropic functions of $\rho$. These conditions, and Equation~(\ref{eq:PP}) 
imply that $p_0(\rho)$ and $\epsilon_0(\rho)$ are locally of the type:
\begin{eqnarray}
  p_0(\rho) &=& K_i\rho^{\Gamma_i}
  \nonumber
\\
  \epsilon_0(\rho) &=& \epsilon_i + \frac{K_i}{\Gamma_i-1} \rho^{\Gamma_i-1}
  \label{eq:e0}
\end{eqnarray}
and all coefficients are set once the polytropic index $\Gamma_i$ ($i=0,\ldots,N-1$),
the transition density $\rho_i$ ($i=1,\ldots,N-1$) and $K_0$ are chosen. This 
defines an $N$-piece piecewise polytropic EOS with an ``ideal-fluid'' thermal 
component. 

The specific enthalpy, in turn, is given by: 
\begin{eqnarray}
h 
  &=& \frac{e+p}{\rho} = \frac{(1+\epsilon)\rho + p}{\rho} \\
  &=& 1+(1-\Gamma_i)\epsilon_i + \Gamma_{\rm th}\epsilon_{\rm th} + \Gamma_i  \epsilon_0 (\rho)
\nonumber 
\end{eqnarray}
Since the relativistic speed of sound can be written as
\begin{eqnarray}
c_s^2 
  &=& \frac{1}{h} \left( \frac{dp}{d\rho} \right)_s \nonumber = \frac{1}{h}
  \left[ \left( \frac{\partial p}{\partial \rho}\right)_\epsilon + \frac{d\epsilon}{d\rho} 
  \left( \frac{\partial p}{\partial \epsilon} \right)_\rho \right]  \\
  &=& \frac{1}{h}\left[ \left( \frac{\partial p}{\partial \rho}\right)_\epsilon + \frac{p}{\rho^2}
        \left( \frac{\partial p}{\partial \epsilon} \right)_\rho \right] \text{,}
\label{eq:cs}
\end{eqnarray}
for a piecewise polytropic EOS we have:
\label{eq:csEOS}
\begin{eqnarray}
c_s^2
  &=& \frac{\Gamma_{\rm th} (\Gamma_{\rm th}-1)\epsilon_{\rm th} + \Gamma_i (\Gamma_i - 1)
  (\epsilon_0(\rho) - \epsilon_i) }{1+(1-\Gamma_i)\epsilon_i + \Gamma_{\rm th}\epsilon_{\rm th} 
  + \Gamma_i \epsilon_0 (\rho)}
\nonumber 
\end{eqnarray}
where $\epsilon_0$ is the same as in Equation~(\ref{eq:e0}) and the index $i$ refers to the $i$-th piece.
At the transition point we have imposed the condition that the pressure and energy density are continuous,
but that does nod hold true for the speed of sound that can  be  discontinuous, \textit{i.e.}, in general $(c_s^2)_i(\rho_i) \neq (c_s^2)_{i+1}(\rho_i)$.

For a piecewise polytropic EOS we calculate the adiabatic index ($\Gamma_1$)  
for adiabatic perturbations as:
\begin{eqnarray}
\Gamma_1 
  &=& \frac{e+p}{p} \left(\frac{dp}{de}\right)_s   
   = \left(\frac{d \ln p}{d \ln \rho}\right)_s\nonumber\\
&=& \Gamma_{\rm th}+(\Gamma_i-\Gamma_{\rm th})\frac{K_i\rho^{\Gamma_i}}{p}
\end{eqnarray}
where $\epsilon$ is the specific internal energy, $K_i$ and $\Gamma_i$ are the polytropic 
constant and exponent in the $i$-th piece of the EOS.

The construction presented here has been considered as a 2D equation
of state in \cite{Camelio:2019}.  They chose to assume as
``entropy'' the quantity $s$ defined by $S=\Sigma(s)=
2/(\Gamma_\mathrm{th}-1) \log(s)$ and to fix the value of the constant
$\alpha$ so that it corresponds to having $\epsilon_\mathrm{th}
\propto s^2 \rho^{\Gamma_\mathrm{th}-1}$. Indeed we should expect, as
shown in ~\cite{Camelio:2019}, that non-barotropic stellar
configurations (including binary neutron star merger remnants) may be
subject to convective instabilities.

Another possibility is to assume as ``entropy'' the quantity $s$ defined by
$S=\Sigma(s)= \log(s)$. With this choice the expressions for the main 
thermodynamical variables, as function of $\rho$ and $s$, are:
\begin{eqnarray}
\epsilon 
   &=& \epsilon_0(\rho) + s^{\Gamma_\mathrm{th}-1} \cdot(\alpha\rho)^{\Gamma_\mathrm{th}-1}\\  
p  &=& p_0(\rho) + (\Gamma_\mathrm{th}-1) \cdot \rho  \cdot s^{\Gamma_\mathrm{th}-1}\cdot (\alpha\rho)^{\Gamma_\mathrm{th}-1}\\  
t  &=& (\Gamma_\mathrm{th}-1) \cdot s^{\Gamma_\mathrm{th}-2} \cdot (\alpha\rho)^{\Gamma_\mathrm{th}-1} 
\end{eqnarray}
One should note that this choice is of difficult physical interpretation in the case of 
$\Gamma_\mathrm{th}=2$ since it would imply that we should interpret the 
density as the temperature of the system. The ``entropy'' ($s$) and the ``temperature'' ($t$) in
terms of the internal energy and density are expressed by: 
\begin{eqnarray}
\epsilon_\mathrm{th} &=& (\epsilon -\epsilon_0(\rho)) \\
t         &=& (\Gamma_\mathrm{th}-1)\cdot (\epsilon_\mathrm{th})^{(\Gamma_\mathrm{th}-2)/(\Gamma_\mathrm{th}-1)}  \cdot (\alpha\rho) \\
s         &=& (\epsilon_\mathrm{th})^{1/(\Gamma_\mathrm{th}-1)} / (\alpha\rho) \\
t \cdot s &=& (\epsilon_\mathrm{th})
\end{eqnarray}
This concludes the discussion of the thermodynamic proprieties of the EOSs used in the present work.

\end{document}